\documentclass[]{aa}
\usepackage{color}
\usepackage{graphicx}

\newcommand{\be}{\begin{equation} }
\newcommand{\ee}{\end{equation} }

\newcommand{\ltapprox}{\raisebox{-0.5ex}{$\,\stackrel{<}{\scriptstyle\sim}\,$}}

\newcommand{\Teff}{T$_{eff}$}
\newcommand{\FeH}{[Fe/H]}

\begin{document}

\title { STELIB: a library of stellar spectra at R$\sim$2000
\thanks{Based on observations collected with the Jacobus Kaptein Telescope,
(owned and operated jointly by the Particle Physics and Astronomy Research
Council of the United Kingdom, the Nederlandse Organisatie voor Wetenschappelijk
Onderzoek of the Netherlands and the Instituto de Astrofísica de Canarias of 
Spain and located in the Spanish Observatorio del Roque de Los Muchachos on La 
Palma which is operated by the Instituto de Astrofísica de Canarias), the 2.3m 
telescope of the Australian National University at Siding Spring, Australia, and 
the VLT-UT1 Antu Telescope (ESO).}
}

\author{J.-F. Le Borgne\inst{1},
G. Bruzual\inst{2},
R. Pell\'o\inst{1},
A. Lan\c con\inst{3},
B. Rocca-Volmerange\inst{4},
B. Sanahuja\inst{5},
D. Schaerer\inst{1},
C. Soubiran\inst{6},
R. V\' \i lchez-G\'omez\inst{7}
 }

\offprints{J.-F. Le Borgne, leborgne@ast.obs-mip.fr}

\institute{Laboratoire d'Astrophysique (UMR 5572),
Observatoire Midi-Pyr\'en\'ees,
14 Avenue E. Belin, F-31400 Toulouse, France
\and
Centro de Investigaciones de Astronom\'\i a, AP 264, 5101-A M\'erida, Venezuela
\and
Observatoire de Strasbourg (UMR 7550), 11 rue de l'Universit\'e, F-67000 
Strasbourg, France
\and
Institut d'Astrophysique de Paris (UMR 7095), 98 bis Boulevard Arago,
F-75014 Paris, France
\and
Departament d'Astronomia i Meteorologia, Universitat de Barcelona,
Mart{\'\i} i Franqu\`es 1, E-08028 Barcelona, Spain
\and
Observatoire de Bordeaux, (UMR 5804), BP 89, F-33270 Floirac, France
\and
Departamento de F\'\i sica, Universidad de Extremadura, Avda. de la Universidad, 
s/n E-10071 C\'aceres, Spain
}

\date{re-submitted version 10 February 2003 }
\authorrunning{J.-F. Le Borgne et al.}{ }
\titlerunning{STELIB, a new library of stellar spectra}{ }

\abstract{ We present STELIB, a new spectroscopic stellar library, available at 
http://webast.ast.obs-mip.fr/stelib. STELIB consists of an homogeneous library
of 249 stellar spectra in the visible range (3200 to 9500\AA), with an 
intermediate spectral resolution (\ltapprox 3\AA) and sampling (1\AA). This 
library includes stars of various spectral types and luminosity classes, spanning 
a relatively wide range in metallicity. The spectral resolution, wavelength and 
spectral type coverage of this library represents a substantial improvement over 
previous libraries used in population synthesis models. The overall absolute 
photometric uncertainty is 3\%.
 }

\maketitle

\section{Introduction}{\label{sect1}}
Evolutionary population synthesis models that describe the chemical and spectral 
evolution of stellar systems in detail are fundamental tools in the analysis of 
observations of both nearby and distant galaxies (e.g. Guiderdoni \&  
Rocca-Volmerange \cite{Guiderdoni}, Buzzoni \cite{Buzzoni}, Bruzual \& Charlot 
\cite{Bruzual}, Fioc \& Rocca-Volmerange \cite{Fioc}). They are needed to 
determine the stellar populations in a variety of systems, spanning a wide range 
of metallicities, from early type galaxies and spiral bulges to star forming 
galaxies at different redshifts. \\
The possibility of building detailed spectro-chemical evolution models of stellar 
populations using evolutionary synthesis techniques is limited by the lack of 
comprehensive empirical libraries of stellar spectra, comprising stars with 
metallicities ranging from well below solar ([Fe/H] from -2 to -1) to above solar 
([Fe/H]$>$0). Direct inversions of galaxy spectra (Pelat \cite{Pelat}, Boisson 
et al. \cite{Boisson}) are also handicaped by this shortage. Current synthesis 
models based on empirical stellar data are mostly restricted to solar metallicity. 
In the visible range, they are largely based on the spectral atlas of Gunn \& 
Stryker (\cite{Gunn}) or the more recent (and not completely independent) atlas 
of Pickles (\cite{Pickles}). \\
The use of theoretical stellar spectra such as Kurucz' (\cite{Kurucz}) instead of 
empirical libraries is {\em a priori} preferable, because they can be computed 
for a dense grid of fundamental parameters (metallicity, gravity, effective 
temperatures), thus avoiding interpolation errors and calibrations. However, 
the resulting synthetic spectra do not in general reproduce the spectral features 
observed in composite stellar populations with the same degree of accuracy as 
models based solely on observed stellar spectra. Methods to achieve photometric 
compatibility between models and data have been developed (Lejeune et al. 
\cite{Lejeune97}), and extended and homogeneous libraries of theoretical spectra 
covering the bulk of the HR-diagram and a wide range of metallicities are now 
available (Lejeune et al. \cite{Lejeune98}, Westera et al. \cite{Westera}). While 
this represents a major improvement, such libraries still suffer from the limited 
resolution ($\sim$20\AA\ in the optical). The determination of stellar 
populations in galaxies up to z$\sim$1 through optical spectroscopy requires 
spectral synthesis capabilities over a broad wavelength range ($\sim$3000\AA\
to $\sim$1,$\mu$m). A minimum spectral resolution of a few \AA\ is necessary to 
obtain constraints on age, metallicity and global stellar kinematics from 
absorption lines. The libraries presently available with a suitable spectral 
resolution (1-3\AA) are often limited to a narrow wavelength range (Jones 
\cite{Jones}, Cenarro et al. \cite{Cenarro}) or are restricted to particular
spectral types (Montes et al. \cite{Montes}). \\
The main objective of our stellar library STELIB is to provide a homogeneous set 
of stellar spectra in the visible range (3200 to 9500\AA), with a relatively 
high spectral resolution (\ltapprox 3\AA) and sampling (1\AA). This library 
includes stars of most spectral types and luminosity classes and spans a 
relatively wide range in metallicity. Most of the stars in our sample have 
measured metallicities. \\
The outline of the paper is the following. In Section \ref{sect2} we present the 
observations. Section \ref{sect3}  describes the selection criteria and the
overall characteristics of the STELIB sample of stars. The data reduction
process is summarized in Section \ref{sect4}. Section \ref{sect5} presents the
content of the library STELIB, presently available through the web. In Section
\ref{sect6} we show some particular applications of STELIB to population 
synthesis studies, and we compare the performances of this library to previous
results. The conclusions of this paper are given in Section \ref{sect7}.

\section{Observations}{\label{sect2}}
The data were obtained during two runs, one at the 1m Jacobus Kaptein Telescope 
(JKT), Roque de los Muchachos Observatory, La Palma, Canary Islands, Spain, 
between 1994 March 28 and April 4, and a second one at the 2.3m of the Australian 
National University at Siding Spring (SSO), Australia, between 1994 December 25 
and 31. On JKT, we used the Richardson-Brealey Spectrograph with the 600 lines/mm
grating. The detector was a EEV7 1242$\times$1152 CCD with a 22.5$\mu$m pixel.
The slit width was 1.5 arcsec. This configuration gives a dispersion of
1.7\AA/pixel and a resolution of about 3\AA\ FWHM. We made use of both blue and 
red optics. With the blue optics, spectra were alternatively obtained with 2 
grating angle settings: 18\degr giving a wavelength range of 2900\AA-5100\AA\ on 
the CCD (useful data start at $\sim$3200\AA\ because of atmospheric cutoff) and 
21\degr giving the wavelength range 4300\AA-6500\AA. With the red optics the 
grating angle settings were 24\degr and 27\degr for the wavelength ranges 
6000\AA-8200\AA\ and 7600\AA-9900\AA, respectively. To maximize the efficiency, 
and to improve the calibration, each night was devoted to a single grating angle 
setting: changing the grating angle was done manually by opening the 
spectrograph. March 29 was an exception because 2 settings with the red optics 
were used (see Table~\ref{tab1} for details). Again to save time, the 
spectrograph was not rotated to align the slit on the paralactic angle, since it 
should have to be done manually on the telescope for each pointing. This should 
have no consequence because of the relatively short wavelength range of each 
individual spectra, the slit width of 1.5 arcsec, and also because we observed 
as close to the meridian as possible (the slit was set vertical when at 
meridian). During the JKT run about 1000 spectra were obtained on about 200 stars.

\begin{table}
\caption[]{\label{tab1} JKT observations: grating angle settings
 }
\begin{flushleft}
\begin{tabular}{rll}
\hline\noalign{\smallskip}
 night & grating   & wavelength range   \\
 1994  & angle     &                    \\
\noalign{\smallskip}
\hline\noalign{\smallskip}
  March 28 & 24\degr & 6000\AA-8200\AA   \\
      March 29 & 24\degr+27\degr & 6000\AA-8200\AA+7600\AA-9900\AA   \\
      March 30 & 21\degr & 4300\AA-6500\AA   \\
      March 31 & 18\degr & 3200\AA-5100\AA   \\
      April 1  & 21\degr & 4300\AA-6500\AA   \\
      April 2  & 24\degr & 6000\AA-8200\AA   \\
      April 3  & 27\degr & 7600\AA-9900\AA   \\
      April 4  & 18\degr & 3200\AA-5100\AA   \\
\noalign{\smallskip}
\hline
\end{tabular}
\end{flushleft}
\end{table}

The spectrograph used at the Siding Spring 2.3m telescope was the Double Beam
Spectrograph.  This instrument has two beams split by a dichroic slide. The 
detectors were 2 1024$\times$1024 CCD's, the blue channel CCD is UV coated. The 
grating used was also a 600 lines/mm giving a dispersion of 1.1\AA/pixel (15 
$\mu$m pixels). The slit width was 2 arcsec on the sky. The spectral resolution 
was less than 3 pixels FWHM with good focus, so about 3\AA. The mode ``vertical 
slit on sky" was used. Three configurations were defined:
\begin{itemize}
\item Configuration 1: wavelength range on
blue channel: 3500-4500\AA;
on red channel: 6470-7550\AA.
\item Configuration 2: wavelength range on
blue channel: 4475-5550\AA;
on red channel: 7500-8570\AA.
\item Configuration 3: wavelength range on
blue channel: 5510-6550\AA;
on red channel: 8530-9650\AA.
\end{itemize}
Table~\ref{tab2} summarizes the configurations used during this run. 36 stars 
were obtained at Siding Spring.  They were selected mainly in the Large 
Magellanic Cloud or among local metal poor stars, in order to improve the 
coverage of stellar parameter space. Some Wolf Rayet stars were also observed.

\begin{table}
\caption[]{\label{tab2} SSO 2.3m spectrograph configurations (see text) }
\begin{flushleft}
\begin{tabular}{rccc}
\hline\noalign{\smallskip}
 night 1994 & config.& \multicolumn{2}{c}{wavelength ranges}    \\
            & number &    blue channel   &   red channel        \\
\noalign{\smallskip}
\hline\noalign{\smallskip}
      December 25 &  1 & 3500\AA-4500\AA & 6470\AA-7550\AA  \\
      December 26 &  2 & 4475\AA-5550\AA & 7500\AA-8570\AA  \\
      December 27 &  3 & 5510\AA-6550\AA & 8530\AA-9650\AA  \\
      December 28 &  1 & 3500\AA-4500\AA & 6470\AA-7550\AA  \\
      December 29 &  2 & 4475\AA-5550\AA & 7500\AA-8570\AA  \\
      December 30 &  3 & 5510\AA-6550\AA & 8530\AA-9650\AA  \\
      December 31 &  1 & 3500\AA-4500\AA & 6470\AA-7550\AA  \\
\noalign{\smallskip}
\hline
\end{tabular}
\end{flushleft}
\end{table}

\begin{figure}
\includegraphics[width=5cm,angle=270]{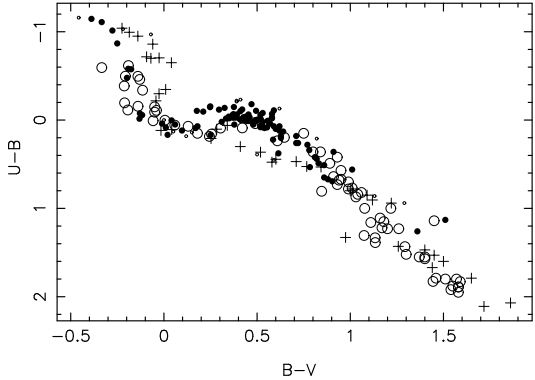}  % {fig01.ps}}
\caption{ U-B vs B-V for the stars in STELIB corrected for interstellar
extinction.
The different symbols represent different stellar spectral classes:
full circles are dwarf main sequence stars (class V), open circles, giants
(class III) and plus sign, super-giants of classes I and II.
Small circles are used for stars with no spectral class determination.
} \label{fig1}
\end{figure}

\begin{figure} [h]
\begin{tabular}{l}
\includegraphics[width=5cm,angle=270]{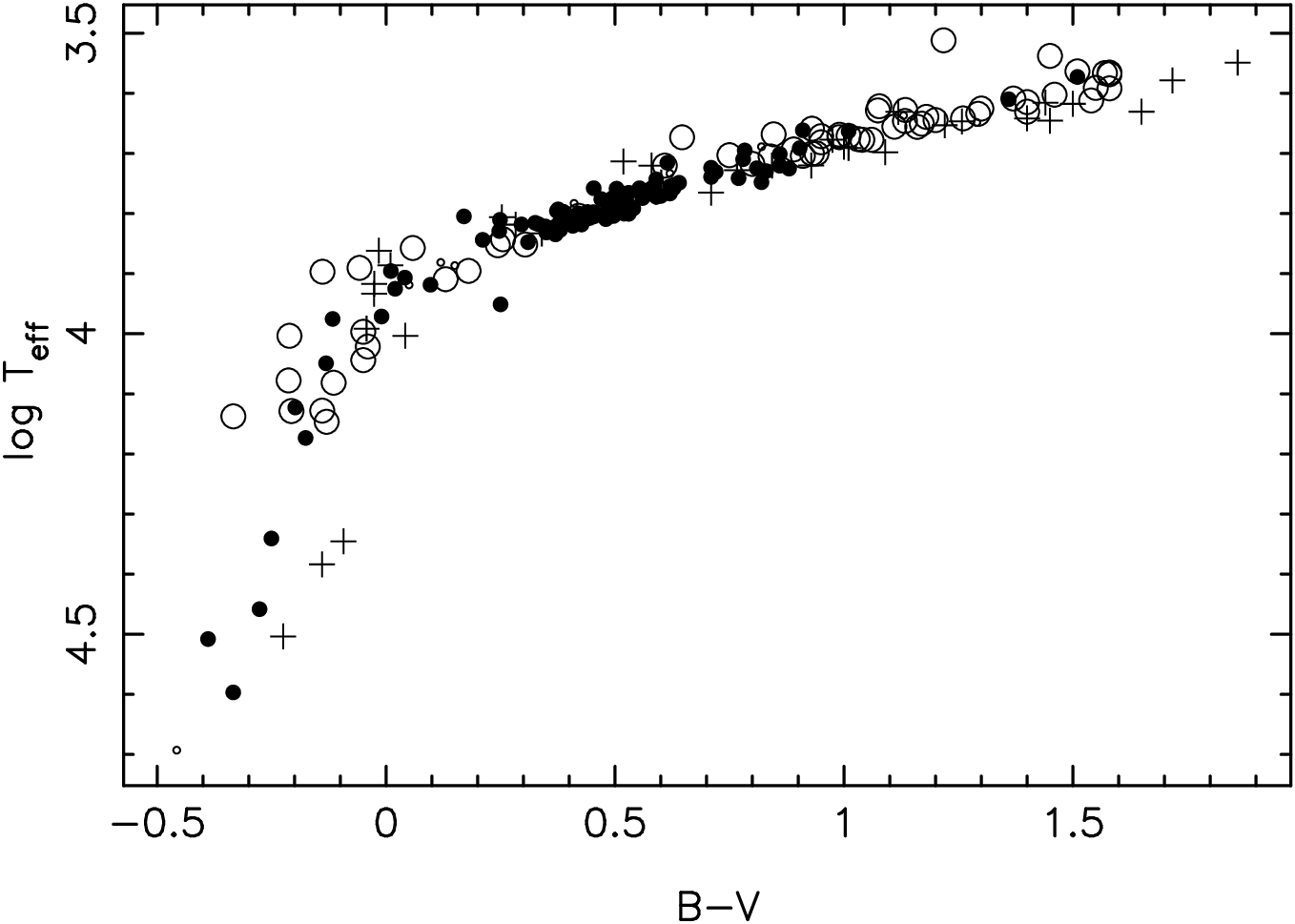}  % {fig03.ps}}
\\
\includegraphics[width=5cm,angle=270]{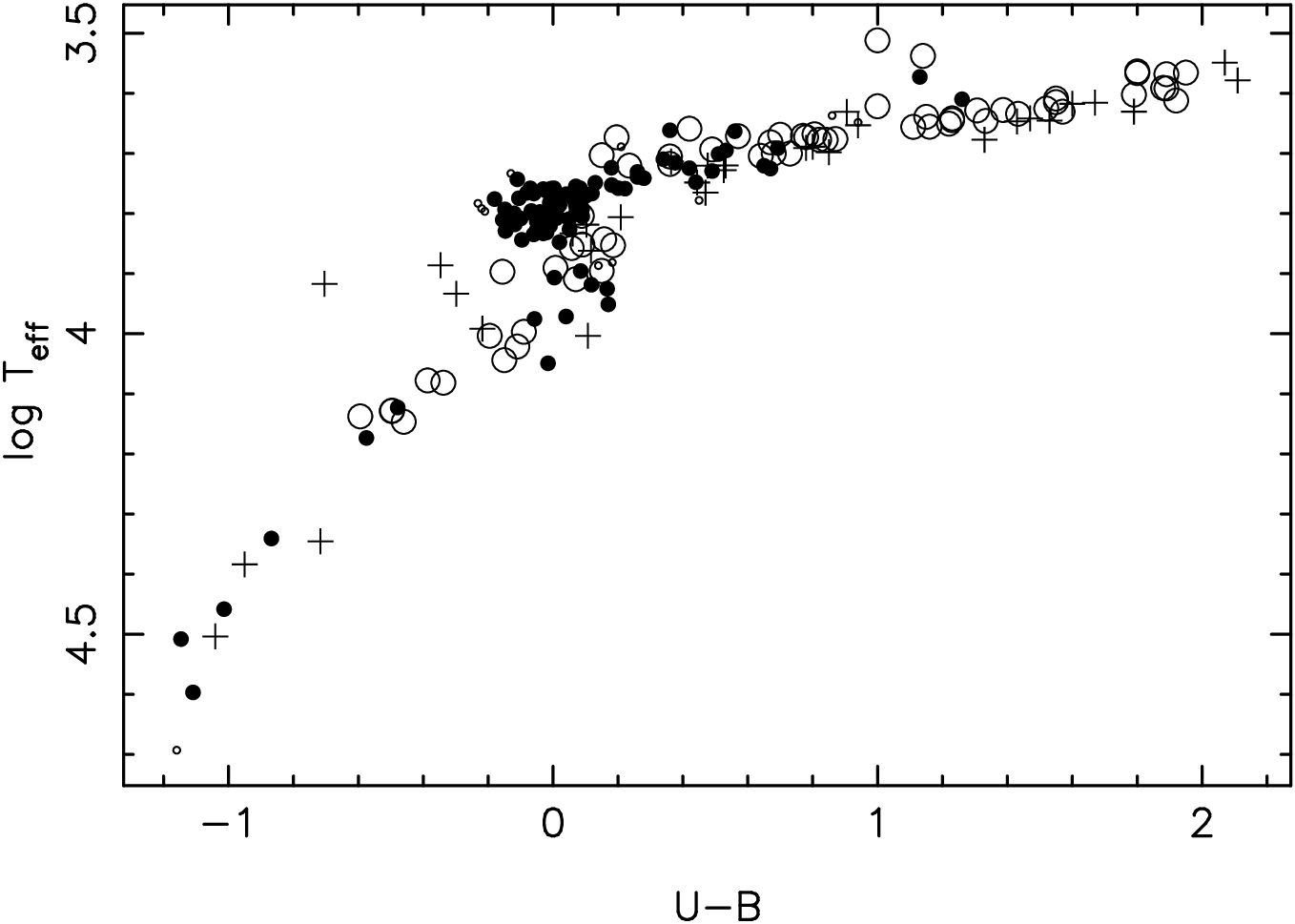}  % {fig04.ps}}
\\
\includegraphics[width=5cm,angle=270]{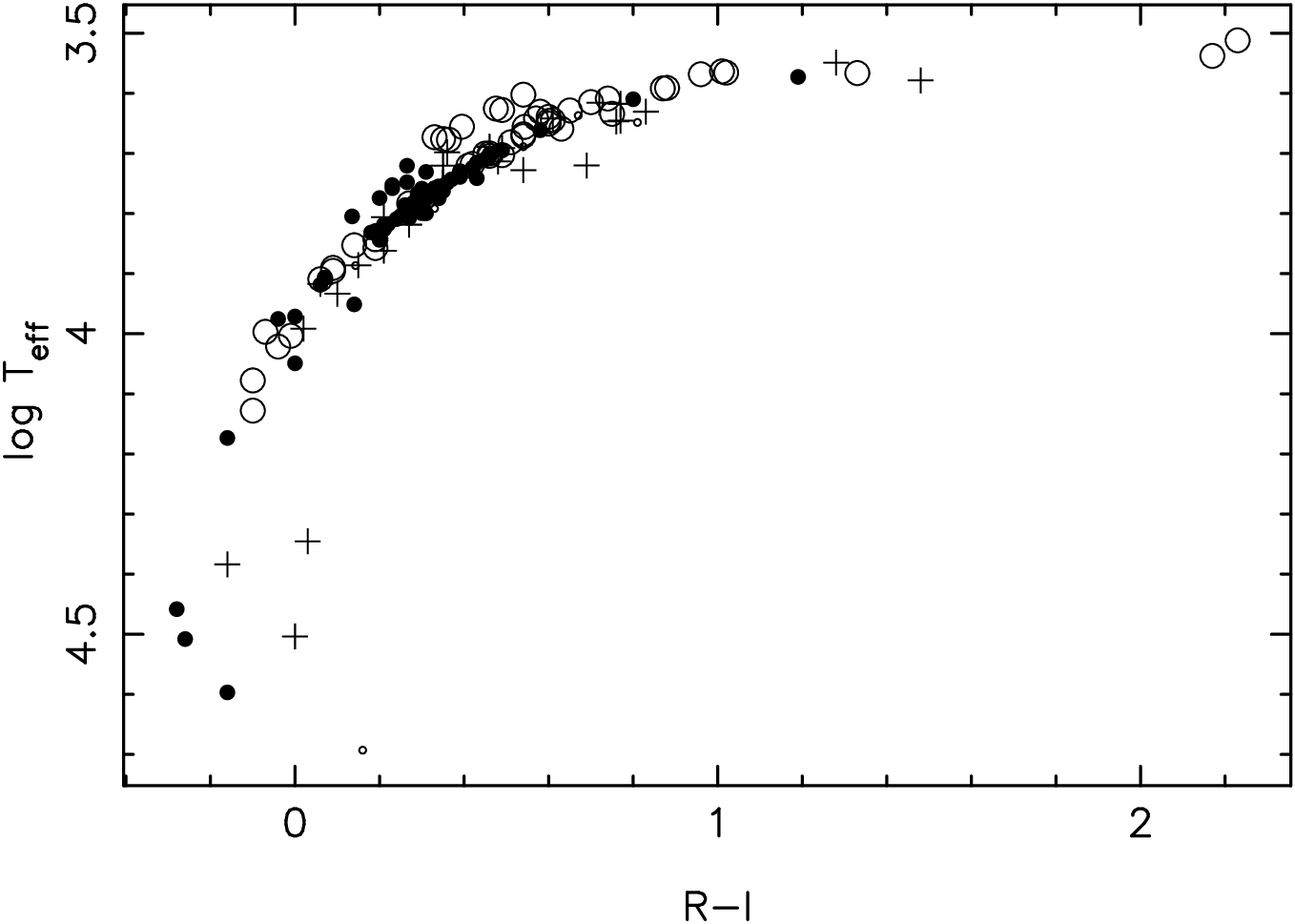}  % {fig05.ps}}
\\
\end{tabular}
\caption[]{\Teff vs B-V, U-B and R-I. The colors are corrected for interstellar
extinction. Symbols are the same as in Fig.~\ref{fig1}. } \label{fig2}
\end{figure}

\begin{figure}
\begin{tabular}{l}
\includegraphics[width=5cm,angle=270]{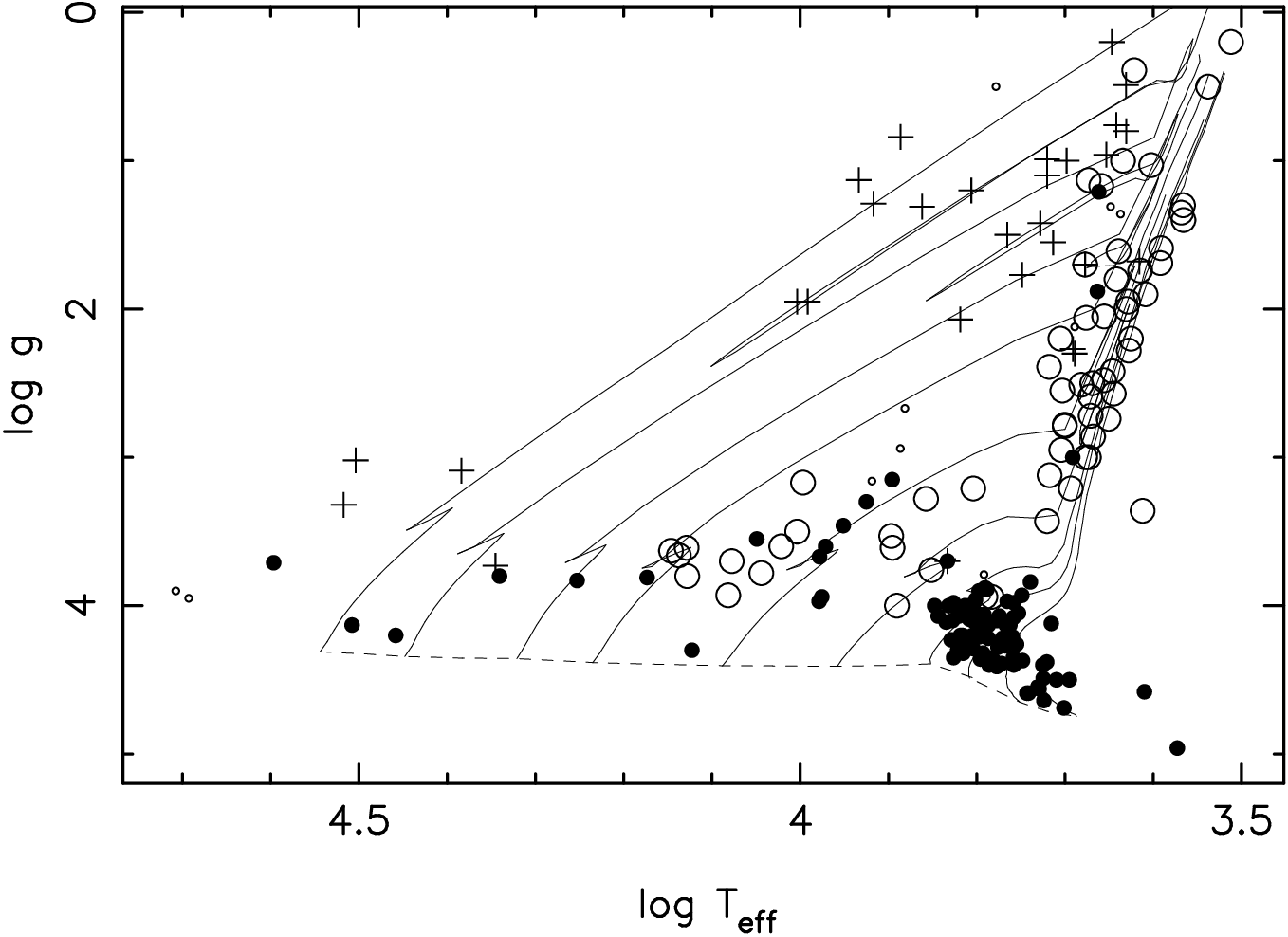}  % {fig02.ps}}
\\
\includegraphics[width=5cm,angle=270]{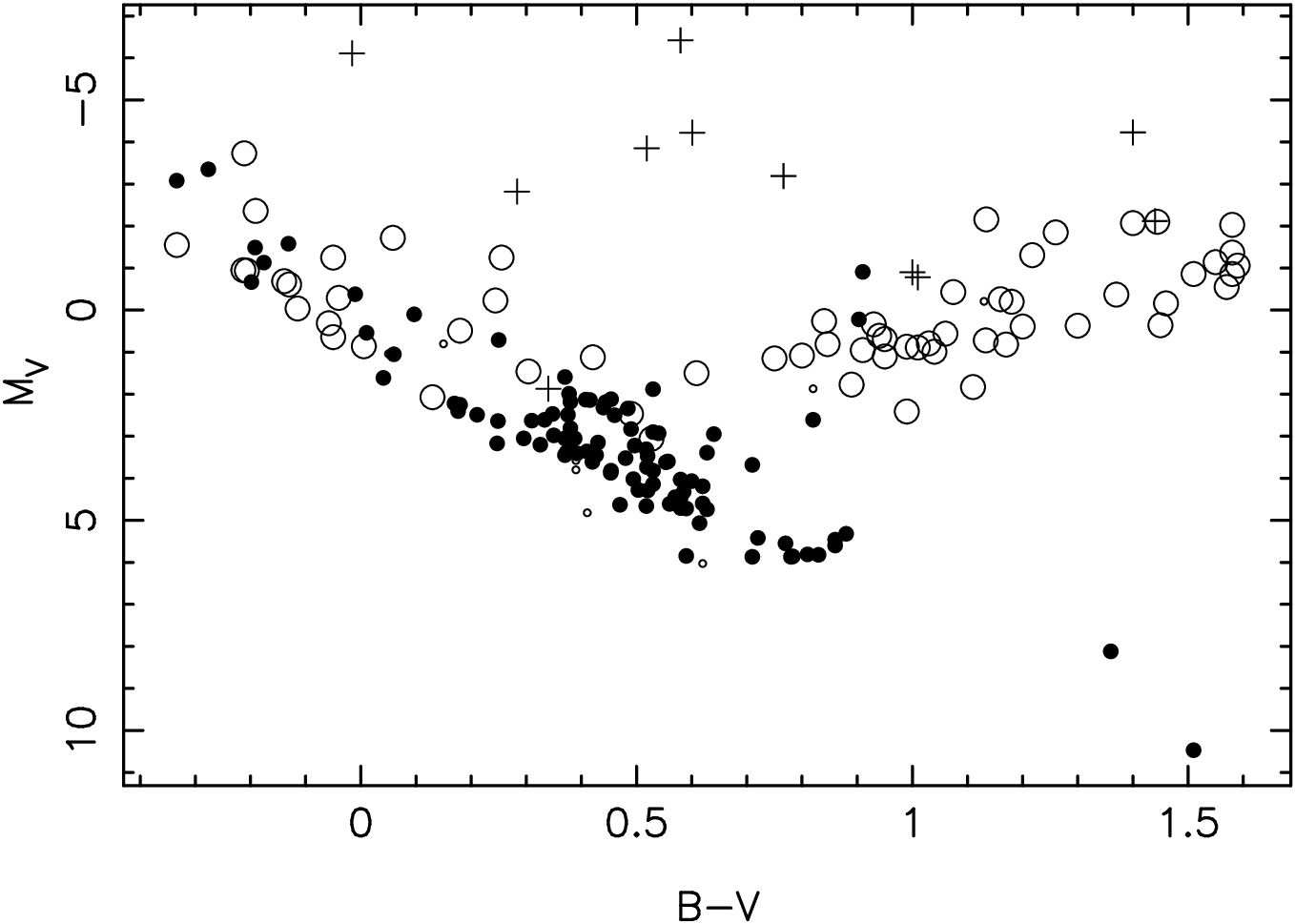}  % {fig06.ps}}
\\
\includegraphics[width=5cm,angle=270]{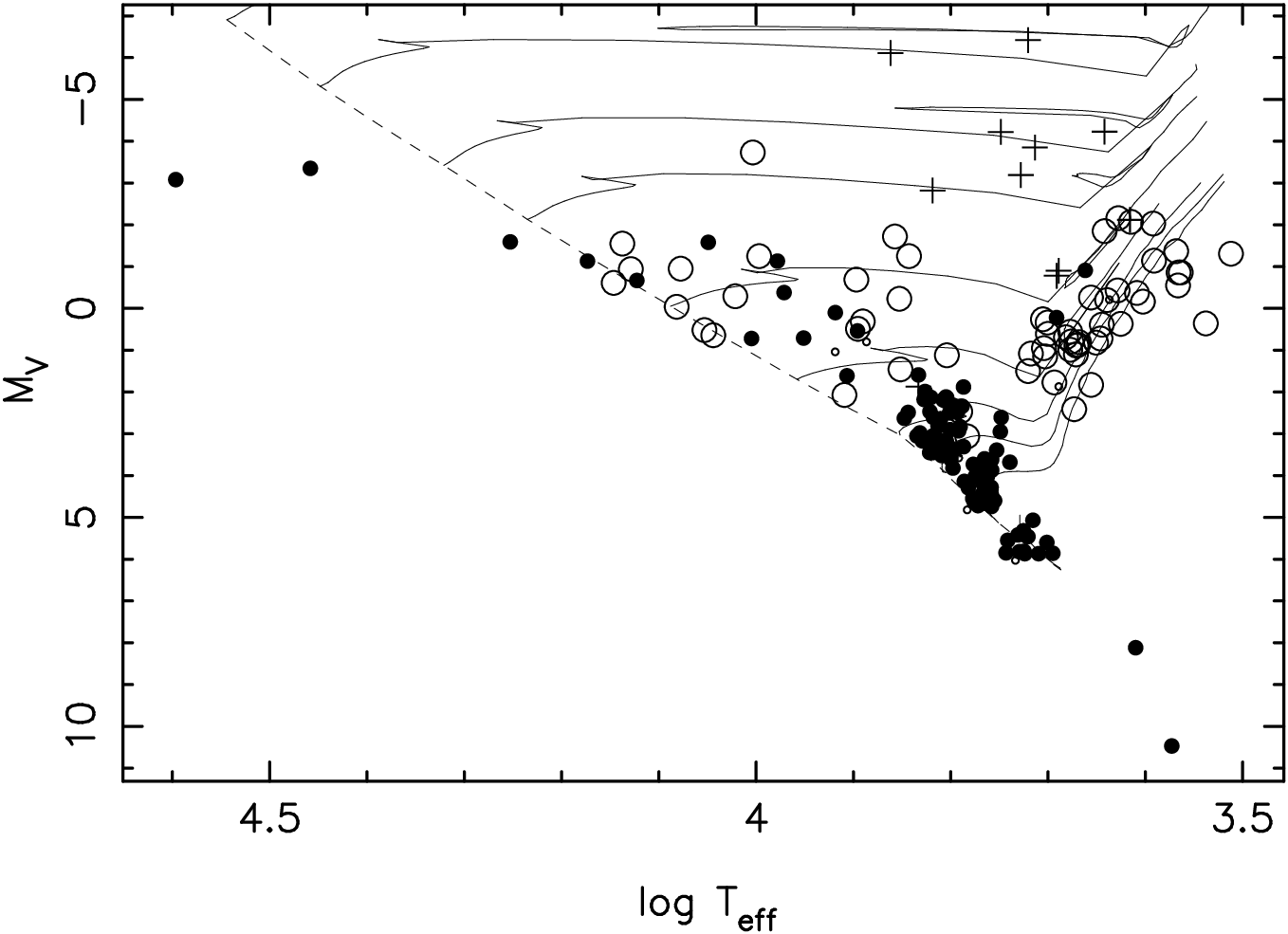}  % {fig07.ps}}
\\
\end{tabular}
\caption[]{ Log(g) vs log(\Teff) and  HR diagrams.
Symbols are the same as in Fig.~\ref{fig1}.} \label{fig3}
\end{figure}

\begin{figure}
\includegraphics[width=5cm,angle=270]{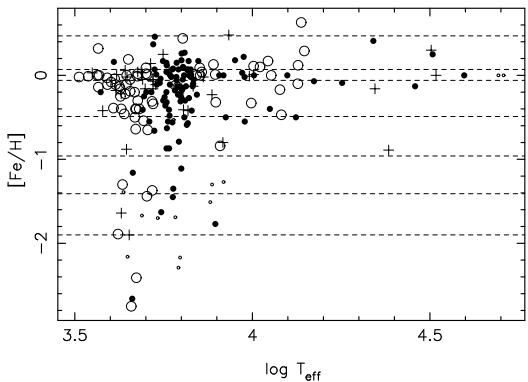}  % {fig08.ps}}
\caption{ Metallicity distribution.  Symbols are the same as in Fig.~\ref{fig1}.
Horizontal lines show the limits of the subsets described in Table~\ref{tab5} } \label{fig4}
\end{figure}

\section{Star selection}{\label{sect3}}
Most stars were originally selected from the catalogue of Cayrel de Strobel et 
al.(\cite{Cayrel92}) according to the value of \FeH. Additional samples of 62 and 
45 stars were selected to include targets with either near-IR spectra (from 
Lan\c con \& Rocca-Volmerange \cite{Lancon92}) and/or UV data (from IUE) 
respectively. \\
The Tables~\ref{cat1} to \ref{cat6} give the list of the 249 stars included in 
the library. Most of the atmospheric parameters (T$_{eff}$, log(g), \FeH) listed 
in Tables~\ref{cat1} to \ref{cat6} come from the 2 latest editions of the 
Catalogue of [Fe/H] determinations (Cayrel de Strobel et al. \cite{Cayrel97}, 
\cite{Cayrel01}). This compilation was complemented by accurate T$_{eff}$ listed
in Blackwell \& Lynas-Gray (\cite{Blackwell}), di Benedetto (\cite{diBenedetto})
and Alonso et al. (\cite{Alonso96}, \cite{Alonso99}). We have also used the V-K 
colour index, when available, calibrated into T$_{eff}$ using the formulae of 
Alonso et al. (\cite{Alonso96}, \cite{Alonso99}). Multiple determinations of 
atmospheric parameters for the same star were averaged, giving more weight to the 
most recent ones. Several stars with unknown atmospheric parameters were also 
part of the ELODIE database (Prugniel \& Soubiran \cite{Prugniel}). In that case 
we give the parameters determined by the TGMET method (Katz et al. \cite{Katz}).
\\
Absolute magnitudes M$_{\textrm v}$ were derived from the Hipparcos parallax and
TYCHO2 V$_T$ apparent magnitude, transformed into V Johnson band (H{\o}g et al.
\cite{Hog}) and corrected with Av measured on the spectra. M$_{\textrm v}$ is 
only given for stars having a relative parallax error lower than 30\%. 
Uncertainties correspond to one $\sigma$ errors on parallaxes and V magnitudes.\\
Most of the stars in the library have accurate UBV photometry available from the
Lausanne "General Catalogue of Photometric Data" compiled by Mermilliod et al.
(\cite{Mermilliod}) and about half of them have R and I photometry. 
Fig.~\ref{fig1} shows the U-B vs B-V diagram corrected for the interstellar
extinction with A$_V$/E$_{B-V}$=3.1, E$_{U-V}$/E$_{B-V}$=1.59,
E$_{R-V}$/E$_{B-V}$=-0.88 and E$_{I-V}$/E$_{B-V}$ = -1.60. The relations
\Teff\ versus color indices are displayed in Fig.~\ref{fig2}. Finally, HR-type 
diagrams are shown in Fig.~\ref{fig3}. In the log(g)/\Teff\ diagram and 
M$_V$/\Teff\ diagrams, evolutionary tracks from the Geneva models (Schaller et 
al. \cite{Schaller}) are displayed for solar metallicity. In the diagram M$_V$ 
versus \Teff\ and M$_V$ versus B-V, M$_V$ are from the Hipparcos catalog 
(Perryman et al., \cite{Perryman}). Fig.~\ref{fig4} shows the distribution of 
\FeH\ as a function of \Teff.

\section{Data reduction}{\label{sect4}}
The basic data reduction was performed with iraf\footnote{IRAF is distributed by 
the National Optical Astronomy Observatories, USA, which are operated by the 
Association of Universities for Research in Astronomy, Inc., under cooperative 
agreement with the National Science Foundation, USA.}
except for the flux calibration of JKT data which appeared to demand non-standard 
procedures. \\
The wavelength calibration was done thanks to the acquisition of arc spectra from 
a Cu-Ne lamp for the JKT data and from He-Ar, Ar-Ne and Cu-Ar lamps for the SSO 
data. The typical number of lines used was 30 to 50. The rms of the residuals is 
of the order of 0.1 \AA.

\begin{table}
\centering
\caption[]{\label{tab3} Standard stars observed at JKT}
\begin{flushleft}
\begin{tabular}{lccl}
\hline\noalign{\smallskip}
 star &  num. of       &  rms   & reference  \\
      &  spectra       &        &            \\
\noalign{\smallskip}
\hline\noalign{\smallskip}
\object{HR2422}                & 19 &  0.058   &  Whiteoak (\cite{Whiteoak}) \\
\object{HR3454}, $\eta$ Hya    & 31 &  0.024   &  Hamuy et al., \cite{Hamuy92},  \\
                               &    &          &  \cite{Hamuy94}  \\
\object{HR4963}, $\theta$ Vir  & 29 &  0.020   &  Hamuy et al., \cite{Hamuy92}, \\
                               &    &          &  \cite{Hamuy94}, Hayes \cite{Hayes70} \\
\object{HR3982}, Regulus       & 27 &  0.020   &  Cochran (\cite{Cochran}),  \\
                               &    &          &  Hayes \cite{Hayes70}  \\
\object{HR5511}, 109 Vir       & 17 &  0.036   &  Cochran (\cite{Cochran}),  \\
                               &    &          &  Johnson (\cite{Johnson})  \\
\object{HR7001}, Vega          & 11 &  0.022   &  Hayes, \cite{Hayes}  \\
\object{HR7589}                & 15 &  0.041   &  Whiteoak (\cite{Whiteoak}) \\
\noalign{\smallskip}
\hline
\end{tabular}
\end{flushleft}
\end{table}

Table~\ref{tab3} gives the list of the standard stars observed at JKT. In 
average, 18 spectra of standard stars were obtained each night, enough to allow 
checking for atmospheric extinction. The examination of these spectra revealed 
strongly varying atmospheric extinction during the observations. Our 
interpretation is that the strong wind blowing from east was carrying dust from 
the Sahara desert. But we cannot exclude that it comes from differential 
atmospheric loss in the JKT narrow slit. \\
As a consequence, the direct use of the standard stars spectra, with a standard
procedure to flux calibrate the spectra was not feasible. We then built a 
procedure to take into account various factors which affect the atmospheric 
extinction both in its absolute value and its dependence with wavelength. \\
The normal atmospheric extinction is modeled by the mean atmospheric extinction 
curve versus wavelength and the airmass at time of observation. The "abnormal" 
extinction, possibly due to dust, is likely to change rapidly during one night. 
We calibrated this effect by using any observed star as a photometric standard 
star. The UBVRI photometry of most of our program stars are available in the 
Lausanne database (http://obswww.unige.ch/gcpd/gcpd.html) (Mermilliod et al., 
\cite{Mermilliod}). However to do this, it has been necessary to take the 
variation of seeing into account. The seeing, measured on each spectrum from the 
profile of the star image along the slit, appeared to change significantly during 
the nights, typically between 0.6\arcsec and 1.5\arcsec. The light lost outside 
the slit differs as the seeing varies. To model it, we took into account the 
stellar profile and the slit width. The detailed modelling process was performed 
individually for each star. This operation gives an absolute mean value over the 
wavelength for a given grating angle setting to scale the spectra. Then, the 
observations of spectrophotometric standard stars were used to analyse the 
wavelength dependence of the additionnal extinction.

\begin{figure}
\includegraphics[width=5cm,angle=270]{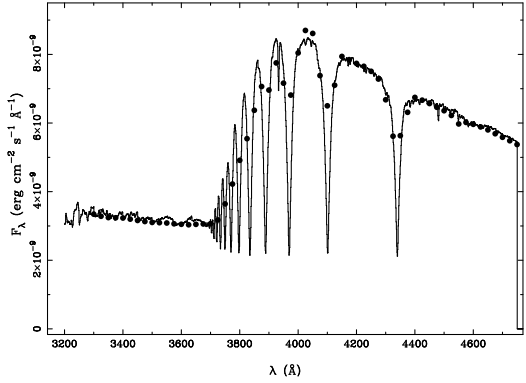}
\caption{ Comparison of Vega spectrum in the UV with published SED (filled circles). }
\label{fig5}
\end{figure}

\begin{figure}
\includegraphics[width=6.5cm,angle=270]{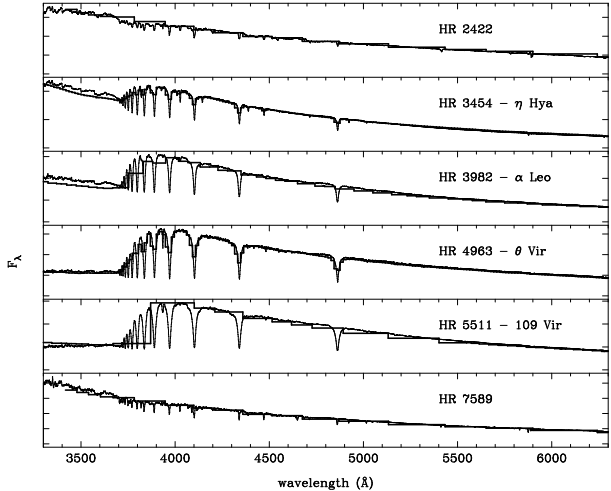}
\caption{ Comparison of calibrated standard stars spectra observed at JKT
with published SED (bold curve). }
\label{fig6}
\end{figure}

\begin{figure}
\includegraphics[width=6cm,angle=270]{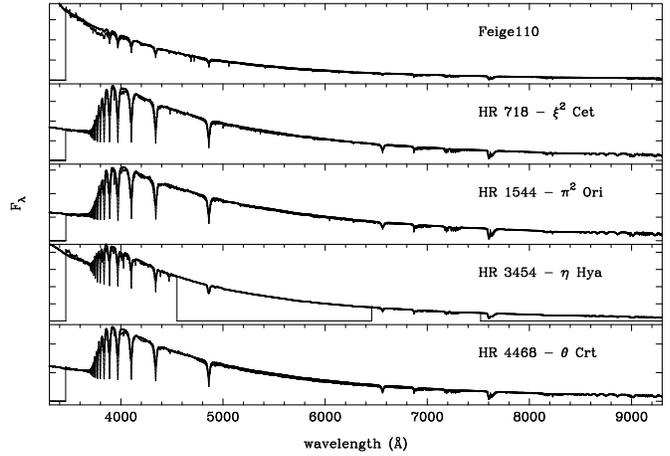}
\caption{ Comparison of calibrated standard stars spectra observed at SSO 2.3m
with published SED (bold curve). }
\label{fig7}
\end{figure}

Fig.~\ref{fig5} and \ref{fig6} show the comparison of the calibrated standard 
stars spectra with published SED. The spectrum of Vega (Fig.~\ref{fig5}) was 
obtained only in the shortest wavelenth setting because of its brightness. The 
comparison for the other standards are shown in Fig.~\ref{fig6}. The rms of 
difference between calibrated spectra of standard stars and published spectra, 
expressed in magnitude, are given in Table~\ref{tab3}. They are computed avoiding 
the strong absorption lines where the difference of wavelength sampling 
introduces large dispersions.These rms are between 0.02 and 0.04 magnitude.
An exception is HR2422 for which the rms is 0.058. \\
The SSO spectra were reduced using the standard procedures for flux calibration.
Table~\ref{tab4} lists the standard stars used. Two target stars and one standard 
star observed in this run were also observed at JKT. We used these stars as an 
additional check on the reliability of the complex flux calibration procedure 
applied to JKT data. A good agreement was obtained between the two independent 
set of spectra. Fig.~\ref{fig7} shows the comparison of the calibrated standard 
stars spectra with published SED. As for the JKT standard stars, the rms of 
difference between calibrated spectra of standard stars and published spectra, 
are given in Table~\ref{tab4}. They are also computed avoiding the strong 
absorption lines. The rms have similar values between 0.02 and 0.04 magnitude. 
One standard star, $\eta$ Hya, has been observed during both runs. The rms of the 
difference is 0.031 magnitude, of the same order than the rms of the difference 
between observed and published spectra. Thus, we can consider that 0.03 magnitude 
is the typical absolute photometric uncertainty of the library. In addition, the 
detailed comparison between the synthetic photometry derived from the STELIB
library and the Lausanne database is presented and discused in Appendix A. 
Tables \ref{antab1} to \ref{antab4} provide with the UVBRI synthetic photometry 
for STELIB stars.

\begin{table}
\centering
\caption[]{\label{tab4} Standard stars observed at SSO}
\begin{flushleft}
\begin{tabular}{lcl}
\hline\noalign{\smallskip}
 star &  rms   &   reference  \\
      & (mags) &              \\
\noalign{\smallskip}
\hline\noalign{\smallskip}
\object{HR 3454}, $\eta$ Hya   &  0.024  &   Hamuy et al., \cite{Hamuy92}, \cite{Hamuy94}  \\
\object{HR 718}, $\xi^2$ Cet   &  0.027  &   Hamuy et al., \cite{Hamuy92}, \cite{Hamuy94}  \\
\object{Feige 110}             &  0.043  &   Oke,\cite{Oke} , Hamuy et al.,  \\
                               &         &    \cite{Hamuy92}, \cite{Hamuy94} \\
\object{HR 4468}, $\theta$ Crt &  0.031  &   Hamuy et al., \cite{Hamuy92}, \cite{Hamuy94}  \\
\object{HR 1544}, $\pi^2$ Ori  &  0.024  &   Hamuy et al., \cite{Hamuy92}, \cite{Hamuy94}  \\
\noalign{\smallskip}
\hline
\end{tabular}
\end{flushleft}
\end{table}
\begin{table*}
\begin{flushleft}
\begin{tabular}{llllrrrrrl}
\hline\noalign{\smallskip}
 Star        &  RA 2000       &  DEC 2000      &      spectr.      &     Teff   &    log(g)  &   [Fe/H]   &     Mv     &      Av    &     identification  \\
 \noalign{\smallskip}
\hline\noalign{\smallskip}
 HD 2857     &  00:31:53.80   &  -05:15:42.3   &           A2      &     7611   &    2.67    &   -1.51    &                         &    0.25    &   -                  \\
 HD 5820     &  00:59:49.68   &  +06:28:59.6   &        M2III      &            &            &            & $-2.10^{+0.53}_{-0.69}$ &    0.70    &   V* WW Psc          \\          
 HD 6268     &  01:03:18.19   &  -27:52:49.7   &        G0III      &     4712   &    1.13    &   -2.41    &                         &    0.60    &   -                  \\          
 HD 9138     &  01:30:10.94   &  +06:08:38.2   &        K4III      &     4061   &    1.90    &   -0.39    & $-0.37^{+0.19}_{-0.20}$ &    0.00    &   $\mu$ Psc          \\          
 HD 12479    &  02:02:35.08   &  +13:28:36.2   &        M2III      &            &            &            & $-1.06^{+0.51}_{-0.66}$ &    0.00    &   -                  \\          
 HD 15318    &  02:28:09.52   &  +08:27:36.3   &        B9III      &    11300   &            &            & $ 0.52^{+0.21}_{-0.23}$ &    0.08    &   $\xi^2$ Cet HR718   \\         
 HD 18191    &  02:55:48.50   &  +18:19:54.0   &     M6IIIvar      &     3450   &    0.50    &   -0.01    & $ 0.36^{+0.25}_{-0.27}$ &    0.00    &   $\rho^2$ Ari V* RZ Ari \\      
 HD 21581    &  03:28:54.45   &  -00:25:02.4   &           G0      &     4885   &    2.12    &   -1.67    & $ 1.87^{+0.56}_{-0.74}$ &    0.00    &   -                  \\          
 HD 26630    &  04:14:53.86   &  +48:24:33.7   &         G0Ib      &     5345   &    1.42    &   -0.11    & $-3.19^{+0.36}_{-0.42}$ &    0.6     &   $\mu$. Per         \\          
 HD 29574    &  04:38:55.74   &  -13:20:47.7   &   G8/K0IIIw.      &     4183   &    0.39    &   -1.89    &                         &    1.0     &   V* HP Eri          \\          
 HD 30739    &  04:50:36.72   &  +08:54:00.9   &         A1Vn      &            &            &            & $ 0.37^{+0.18}_{-0.19}$ &    0.10    &   $\pi^2$ Ori HR 1544\\          
 HD 268623   &  04:52:11      &  -66:42:08     &         B2Ia      &            &            &            &                         &    0.00    &   -                  \\          
 HD 268749   &  04:53:29      &  -69:24:34     &        B7Iab      &            &            &            &                         &    0.00    &   -                  \\          
 HD 30614    &  04:54:03.01   &  +66:20:33.6   &      O9.5Iae      &    31888   &    3.02    &    0.30    &                         &    0.79    &   $\alpha$ Cam       \\          
 HD 32034    &  04:55:11      &  -67:10:10     &         B9Ia      &     8262   &    1.29    &   -0.80    &                         &    0.39    &   -                  \\          
 HD 268819   &  04:55:32.46   &  -69:57:45.1   &           F6      &            &            &            &                         &    0.00    &   -                  \\          
 HD 33133    &  05:03:08      &  -66:40:42     &           WN      &            &            &            &                         &    0.00    &   LMC FD 12          \\          
 HD 33579    &  05:05:55.51   &  -67:53:10.9   &         A3Ia      &     7694   &    0.84    &   -0.23    &                         &    0.56    &   -                  \\          
 HD 32537    &  05:06:40.66   &  +51:35:53.3   &          F0V      &     6978   &    4.07    &   -0.23    & $ 2.49^{+0.06}_{-0.06}$ &    0.4     &   12 Cam V* BM Cam   \\          
 HD 32923    &  05:07:26.68   &  +18:38:42.0   &          G4V      &     5727   &    3.98    &   -0.20    & $ 3.62^{+0.05}_{-0.05}$ &    0.3     &   104 Tau            \\          
 HD 271163   &  05:18:52      &  -65:41:22     &         B3Ia      &            &            &            &                         &    0.00    &   -                  \\          
 HD 34411    &  05:19:08.08   &  +40:06:02.4   &          G0V      &     5845   &    4.13    &    0.01    & $ 4.19^{+0.04}_{-0.04}$ &    0.00    &   $\lambda$ Aur      \\          
 HD 34816    &  05:19:34.53   &  -13:10:36.4   &       B0.5IV      &    28724   &    4.20    &   -0.13    & $-3.35^{+0.51}_{-0.65}$ &    0.02    &   $\lambda$ Lep      \\          
 HD 271182   &  05:21:01.71   &  -65:48:02.4   &        F8...      &     6000   &    0.50    &   -0.53    &                         &    0.00    &   -                  \\          
 HD 35497    &  05:26:17.50   &  +28:36:28.3   &        B7III      &    13429   &    3.80    &   -0.10    &                         &    0.03    &   $\beta$ Tau        \\          
 HD 269687   &  05:31:25      &  -69:05:35     &          B0e      &            &            &            &                         &    0.00    &   -                  \\          
 HD 269697   &  05:31:38.42   &  -67:28:11.6   &         F5Ia      &            &            &            &                         &    0.00    &   -                  \\          
 HD 269698   &  05:31:42      &  -67:38:06     &          O5e      &            &            &            &                         &    0.00    &   -                  \\          
 HD 36512    &  05:31:55.86   &  -07:18:05.5   &          B0V      &    32200   &    4.13    &    0.25    &                         &    0.4     &   $\upsilon$ Ori     \\          
 HD 36673    &  05:32:43.81   &  -17:49:20.3   &         F0Ib      &     7276   &    1.31    &   -0.02    & $-6.11^{+0.56}_{-0.74}$ &    0.7     &   $\alpha$ Lep       \\          
 HD 37680    &  05:34:18      &  -69:45:00     &           WC      &            &            &            &                         &    0.00    &   FD 46              \\          
 HD 269781   &  05:34:22.47   &  -67:01:23.6   &        A0Iae      &            &            &            &                         &    0.00    &   -                  \\          
 HD 38282    &  05:38:53      &  -69:02:01     &           WN      &            &            &            &                         &    0.00    &   FD 70              \\          
 HD 37828    &  05:40:54.62   &  -11:11:59.8   &           K0      &     4335   &    1.36    &   -1.39    & $-0.21^{+0.44}_{-0.54}$ &    0.00    &   -                  \\          
 HD 37394    &  05:41:20.33   &  +53:28:56.4   &          K1V      &     5194   &    4.12    &   -0.20    & $ 5.07^{+0.04}_{-0.04}$ &    0.7     &   -                  \\          
 HD 38247    &  05:45:11.52   &  +18:42:15.7   &        G8Iab      &     4755   &    1.70    &    0.04    &                         &    2.0     &   -                  \\          
 HD 39587    &  05:54:23.08   &  +20:16:35.1   &          G0V      &     5915   &    4.39    &   -0.02    & $ 4.72^{+0.04}_{-0.04}$ &    0.00    &   $\chi^1$ Ori       \\          
 HD 39801    &  05:55:10.29   &  +07:24:25.3   &         M2Ib      &     3540   &    0.00    &    0.03    &                         &    0.00    &   $\alpha$ Ori  V*   \\          
 HD 39866    &  05:56:33.77   &  +28:56:32.2   &         A2II      &    10080   &    1.95    &    0.01    &                         &    0.80    &   -                  \\          
 HD 39949    &  05:57:05.56   &  +27:19:00.1   &         G2Ib      &     5250   &    1.10    &   -0.16    &                         &    0.5     &   -                  \\
 HD 40111    &  05:57:59.66   &  +25:57:14.1   &         B1Ib      &            &            &            &                         &    0.39    &   139 Tau            \\
 HD 41667    &  06:05:03.64   &  -32:59:39.3   &          G8V      &     4605   &    1.88    &   -1.16    &                         &    0.00    &   -                  \\
 \hline\noalign{\smallskip}
 \end{tabular}
 \end{flushleft}
\caption[]{ \label{cat1} catalogue of the stars}
 \end{table*}
 \begin{table*}
 \begin{flushleft}
 \begin{tabular}{llllrrrrrl}
 \hline\noalign{\smallskip}
 Star  &  RA 2000   &  DEC 2000   &
  spectr. &  Teff  & log(g)&  [Fe/H]&  Mv    &  Av &     identification  \\
 \noalign{\smallskip}
 \hline\noalign{\smallskip}
 HD 41593    &  06:06:40.55   &  +15:32:32.5   &          K0V      &     5305   &    4.49    &    0.07    & $ 5.81^{+0.05}_{-0.05}$ &    0.00    &   V* V1386 Ori       \\
 HD 41636    &  06:08:23.14   &  +41:03:21.0   &        G9III      &     4755   &    1.70    &   -0.30    & $ 0.99^{+0.21}_{-0.23}$ &    0.00    &   -                  \\          
 HD 42454    &  06:12:05.49   &  +29:29:31.8   &         G2Ib      &     5250   &    1.10    &   -0.05    &                         &    1.2     &   -                  \\          
 HD 42543    &  06:12:19.10   &  +22:54:30.7   &      M1Ia-ab      &     3789   &    0.00    &   -0.42    &                         &    1.0     &   6 Gem V* BU Gem    \\          
 HD 43153    &  06:15:25.13   &  +16:08:35.5   &          B7V      &    13263   &    4.30    &   -0.50    & $-0.67^{+0.30}_{-0.34}$ &    0.15    &   72 Ori             \\          
 HD 45829    &  06:30:02.29   &  +07:55:16.0   &        K0Iab      &     4435   &    0.20    &   -0.02    &                         &    1.0     &   -                  \\          
 HD 46223    &  06:32:09.31   &  +04:49:24.7   &          O5e      &    49300   &    3.95    &            &                         &    2.1     &   -                  \\          
 HD 47129    &  06:37:24.04   &  +06:08:07.4   &      O8V+...      &            &            &            &                         &    1.13    &   HR2422             \\          
 HD 47839    &  06:40:58.66   &  +09:53:44.7   &         O7Ve      &    39500   &    3.71    &            & $-3.08^{+0.46}_{-0.58}$ &    0.26    &   15 Mon V* S Mon    \\          
 HD 47731    &  06:41:20.90   &  +28:11:47.9   &         G5Ib      &     4990   &    1.00    &   -0.16    &                         &    0.00    &   25 Gem             \\          
 HD 48329    &  06:43:55.93   &  +25:07:52.2   &         G8Ib      &     4384   &    0.76    &    0.06    & $-4.23^{+0.51}_{-0.65}$ &    0.00    &   $\varepsilon$ Gem  \\          
 HD 48682    &  06:46:44.34   &  +43:34:37.3   &          G0V      &     5727   &            &    0.15    & $ 3.87^{+0.04}_{-0.04}$ &    0.3     &   $\psi^5$ Aur A     \\          
 HD 48682B   &  06:46:46      &  +43:35:03     &          M0V      &            &            &            &                         &    0.00    &   $\psi^5$ Aur B     \\
 HD 49933    &  06:50:49.82   &  -00:32:25.5   &          F2V      &     6538   &    4.32    &   -0.59    & $ 3.20^{+0.07}_{-0.07}$ &    0.2     &   -                  \\          
 HD 50420    &  06:55:14.66   &  +43:54:36.2   &        A9III      &     7200   &    3.28    &    0.04    & $-1.72^{+0.41}_{-0.50}$ &    0.75    &   V* V352 Aur        \\          
 HD 52005    &  07:00:15.82   &  +16:04:44.4   &       K4Iab:      &     4271   &    0.80    &   -0.22    &                         &    0.00    &   41 Gem             \\          
 HD 52973    &  07:04:06.54   &  +20:34:13.1   &      G3Ibvar      &     5604   &    1.77    &    0.25    & $-4.22^{+0.57}_{-0.76}$ &    0.4     &   $\zeta$ Gem V*     \\          
 HD 53929    &  07:07:06.48   &  +04:54:38.2   &      B9.5III      &    14011   &    3.63    &    0.29    & $-0.61^{+0.36}_{-0.42}$ &    0.00    &   -                  \\          
 HD 60778    &  07:36:11.79   &  -00:08:14.9   &          A1V      &     8416   &    3.30    &   -0.50    &                         &    0.28    &   -                  \\          
 HD 61064    &  07:37:16.73   &  -04:06:39.7   &        F6III      &     6367   &    3.21    &    0.44    & $ 1.12^{+0.13}_{-0.14}$ &    0.06    &   25 Mon             \\          
 LHS 235     &  07:40:21      &  -17:24:54     &           DF      &            &            &            &                         &    0.00    &   -                  \\          
 HD 63077    &  07:45:35.18   &  -34:10:35.6   &          G0V      &     5731   &    4.08    &   -0.87    & $ 4.46^{+0.04}_{-0.04}$ &    0.00    &   171 Pup            \\          
 HD 64090    &  07:53:32.64   &  +30:36:34.3   &         sdG2      &     5413   &    4.53    &   -1.70    & $ 6.03^{+0.09}_{-0.09}$ &    0.00    &   -                  \\          
 HD 65583    &  08:00:32.24   &  +29:12:54.7   &          G8V      &     5295   &    4.64    &   -0.66    & $ 5.87^{+0.05}_{-0.05}$ &    0.00    &   -                  \\          
 HD 67767    &  08:10:27.23   &  +25:30:29.4   &         G8IV      &     5598   &    4.37    &    0.17    & $ 2.61^{+0.08}_{-0.09}$ &    0.00    &   $\psi$ Cnc         \\          
 HD 69897    &  08:20:03.87   &  +27:13:07.0   &          F6V      &     6275   &    4.21    &   -0.31    & $ 3.82^{+0.05}_{-0.05}$ &    0.02    &   $\chi$ Cnc         \\          
 HD 70272    &  08:22:50.13   &  +43:11:18.1   &        K5III      &     3900   &    1.59    &   -0.03    & $-1.14^{+0.21}_{-0.23}$ &    0.00    &   31 Lyn V* BN Lyn   \\          
 HD 72184    &  08:32:55.06   &  +38:01:00.4   &        K2III      &     4525   &    2.05    &   -0.05    & $ 1.83^{+0.12}_{-0.13}$ &    0.00    &   -                  \\          
 HD 72324    &  08:33:00.14   &  +24:05:05.7   &        G9III      &     4743   &    2.06    &   -0.01    & $ 0.79^{+0.29}_{-0.34}$ &    0.00    &   $\upsilon^2$ Cnc   \\          
 HD 74280    &  08:43:13.49   &  +03:23:55.2   &          B3V      &    17899   &    3.83    &   -0.09    & $-1.59^{+0.29}_{-0.32}$ &    0.08    &   $\eta$ Hya HR3454  \\          
 HD 74739    &  08:46:41.83   &  +28:45:36.0   &       G8Iab:      &     4910   &    2.27    &   -0.05    & $-0.78^{+0.27}_{-0.30}$ &    0.00    &   $\iota$ Cnc A      \\          
 HD 75732    &  08:52:36.13   &  +28:19:53.0   &          G8V      &     5256   &    4.38    &    0.37    & $ 5.46^{+0.04}_{-0.04}$ &    0.00    &   $\rho$ Cnc A       \\          
 HD 76151    &  08:54:18.19   &  -05:26:04.3   &          G3V      &     5728   &    4.40    &    0.04    & $ 4.74^{+0.05}_{-0.05}$ &    0.1     &   -                  \\          
 HD 76943    &  09:00:38.75   &  +41:47:00.4   &          F5V      &     6575   &    4.03    &    0.07    & $ 2.61^{+0.06}_{-0.06}$ &    0.3     &   10 UMa             \\          
 HD 77350    &  09:02:44.27   &  +24:27:10.6   &        A0III      &    10506   &    3.60    &    0.10    & $-0.29^{+0.29}_{-0.33}$ &    0.00    &   $\nu$ Cnc          \\          
 HD 77729    &  09:05:04.74   &  +26:09:53.4   &        K4III      &     4271   &    2.00    &   -0.40    &                         &    0.00    &   -                  \\          
 HD 78316    &  09:07:44.82   &  +10:40:05.6   &     B8IIImnp      &    13449   &    3.61    &    0.12    & $-0.94^{+0.29}_{-0.33}$ &    0.3     &   kap Cnc V*         \\
 HD 78418    &  09:08:47.42   &  +26:37:48.0   &         G5IV      &     5659   &    4.05    &   -0.26    & $ 3.39^{+0.08}_{-0.09}$ &    0.1     &   75 Cnc             \\
 HD 79158    &  09:13:48.23   &  +43:13:04.5   &     B8IIImnp      &    13727   &    3.66    &    0.63    & $-1.55^{+0.31}_{-0.35}$ &    0.6     &   36 Lyn             \\
 HD 79452    &  09:15:14.36   &  +34:38:00.2   &        G6III      &     5072   &    2.20    &   -0.65    & $ 0.26^{+0.28}_{-0.31}$ &    0.00    &   -                  \\
 HD 80081    &  09:18:50.67   &  +36:48:10.4   &          A1V      &            &            &            & $ 1.05^{+0.08}_{-0.08}$ &    0.00    &   38 Lyn             \\
 HD 81192    &  09:24:45.39   &  +19:47:12.8   &        G7III      &     4692   &    2.59    &   -0.64    & $ 1.10^{+0.30}_{-0.34}$ &    0.00    &   -                  \\
\hline\noalign{\smallskip}
 \end{tabular}
 \end{flushleft}
\caption[]{ \label{cat2} catalogue of the stars (cont'd)}
 \end{table*}
 \begin{table*}
 \begin{flushleft}
 \begin{tabular}{llllrrrrrl}
 \hline\noalign{\smallskip}
 Star  & RA 2000   &  DEC 2000   &
  spectr. &   Teff  &  log(g)&  [Fe/H]&  Mv    &  Av &      identification  \\
 \noalign{\smallskip}
 \hline\noalign{\smallskip}
 HD 81809    &  09:27:46.92   &  -06:04:15.7   &          G2V      &     5610   &    3.93    &   -0.30    & $ 2.95^{+0.09}_{-0.09}$ &    0.00    &   -                  \\
 HD 82210    &  09:34:28.97   &  +69:49:48.6   &     G4III-IV      &     5253   &    3.43    &   -0.34    & $ 1.50^{+0.06}_{-0.06}$ &    0.5     &   -                  \\          
 HD 83632    &  09:40:34.09   &  +26:00:14.4   &        K2III      &     4308   &    1.00    &   -1.30    &                         &    0.3     &   -                  \\          
 HD 84937    &  09:48:55.87   &  +13:44:46.1   &         sdF5      &     6257   &    4.04    &   -2.17    & $ 3.80^{+0.20}_{-0.22}$ &    0.00    &   -                  \\          
 HD 85235    &  09:52:06.36   &  +54:03:51.4   &         A3IV      &    11200   &    3.55    &   -0.40    & $-1.58^{+0.24}_{-0.27}$ &    0.5     &   $\varphi$ UMa      \\          
 HD 86161    &  09:54:52.91   &  -57:43:38.3   &           WN      &            &            &            &                         &    0.00    &   WR 16 V* V396 Car  \\          
 HD 86728    &  10:01:01.02   &  +31:55:29.0   &          G1V      &     5737   &    4.25    &    0.12    & $ 4.32^{+0.04}_{-0.04}$ &    0.2     &   20 LMi             \\          
 HD 86986    &  10:02:29.48   &  +14:33:27.0   &          A1V      &     7867   &    3.15    &   -1.77    & $ 0.54^{+0.51}_{-0.65}$ &    0.34    &   -                  \\          
 HD 87141    &  10:04:36.35   &  +53:53:30.2   &          F5V      &     6369   &    3.99    &    0.09    & $ 2.14^{+0.09}_{-0.09}$ &    0.2     &   -                  \\          
 HD 87737    &  10:07:19.95   &  +16:45:45.6   &         A0Ib      &     9806   &    1.95    &   -0.01    &                         &    0.04    &   $\eta$ Leo         \\          
 HD 87696    &  10:07:25.73   &  +35:14:40.9   &          A7V      &            &            &            & $ 2.25^{+0.07}_{-0.07}$ &    0.00    &   21 LMi             \\          
 HD 87822    &  10:08:15.94   &  +31:36:15.4   &          F4V      &     6615   &    4.07    &    0.17    & $ 2.13^{+0.13}_{-0.13}$ &    0.1     &   -                  \\          
 HD 87901    &  10:08:22.46   &  +11:58:01.9   &          B7V      &    12540   &            &            &                         &    0.08    &   $\alpha$ Leo Regulus HR3982  \\
 HD 88355    &  10:11:38.19   &  +13:21:18.7   &          F7V      &     6423   &            &    0.00    & $ 2.19^{+0.14}_{-0.15}$ &    0.05    &   34 Leo             \\          
 HD 88609    &  10:14:28.98   &  +53:33:39.6   &      G5IIIwe      &     4560   &    1.17    &   -2.75    &                         &    0.00    &   -                  \\          
 HD 89025    &  10:16:41.40   &  +23:25:02.4   &        F0III      &     6965   &            &            & $-1.25^{+0.15}_{-0.16}$ &    0.17    &   $\zeta$ Leo        \\          
 HD 89254    &  10:17:37.90   &  -08:04:08.1   &        F2III      &     7106   &    3.76    &    0.09    & $ 1.46^{+0.11}_{-0.11}$ &    0.05    &   $\varepsilon$ Sex  \\          
 HD 89758    &  10:22:19.80   &  +41:29:58.0   &        M0III      &     3700   &    1.35    &    0.00    & $-1.37^{+0.14}_{-0.14}$ &    0.00    &   $\mu$ UMa          \\          
 HD 90277    &  10:25:54.87   &  +33:47:46.5   &          F0V      &     8937   &    3.46    &    0.18    & $ 0.71^{+0.12}_{-0.13}$ &    0.00    &   30 LMi             \\          
 HD 90508    &  10:28:03.81   &  +48:47:13.4   &          G1V      &     5735   &    4.35    &   -0.32    & $ 4.28^{+0.06}_{-0.06}$ &    0.3     &   -                  \\          
 HD 90537    &  10:27:53.09   &  +36:42:26.9   &     G8III-IV      &     5061   &    2.95    &    0.01    & $ 0.95^{+0.10}_{-0.10}$ &    0.00    &   $\beta$ LMi        \\          
 HD 90839    &  10:30:37.76   &  +55:58:50.2   &          F8V      &     6052   &    4.36    &   -0.18    & $ 4.29^{+0.04}_{-0.04}$ &    0.00    &   36 UMa             \\          
 HD 91316    &  10:32:48.68   &  +09:18:23.7   &         B1Ib      &    24200   &    3.09    &   -0.89    &                         &    0.0     &   $\rho$ Leo V*      \\          
 HD 92740    &  10:41:17.52   &  -59:40:36.9   &       WNs...      &            &            &            &                         &    1.30    &   WR 22 V* V429 Car  \\          
 HD 92809    &  10:41:38.33   &  -58:46:18.8   &           WC      &            &            &            &                         &    0.00    &   WR 23              \\          
 HD 92769    &  10:43:01.95   &  +26:19:32.5   &         A4Vn      &     6380   &    4.10    &   -0.15    & $ 2.22^{+0.09}_{-0.10}$ &    0.00    &   40 LMi             \\          
 HD 93250    &  10:44:45.04   &  -59:33:54.7   &           O5      &    51000   &    3.90    &            &                         &    1.58    &   -                  \\          
 HD 93430    &  10:47:22.18   &  +16:55:20.7   &           G0      &            &            &            & $ 3.90^{+0.30}_{-0.34}$ &    0.00    &   -                  \\          
 HD 93765    &  10:49:53.73   &  +27:58:25.9   &          F5V      &     6811   &    3.70    &   -0.10    & $ 1.59^{+0.14}_{-0.15}$ &    0.00    &   44 LMi             \\          
 HD 94028    &  10:51:28.29   &  +20:16:43.0   &          F4V      &     5969   &    4.28    &   -1.45    & $ 4.63^{+0.15}_{-0.16}$ &    0.0     &   -                  \\          
 HD 94264    &  10:53:18.64   &  +34:12:56.0   &     K0III-IV      &     4657   &    2.86    &   -0.20    & $ 0.81^{+0.07}_{-0.07}$ &    0.6     &   46 LMi             \\          
 HD 94247    &  10:53:34.52   &  +54:35:06.5   &        K3III      &     4243   &    2.28    &   -0.16    & $-2.16^{+0.28}_{-0.32}$ &    0.7     &   44 UMa             \\          
 HD 95345    &  11:00:33.64   &  +03:37:03.1   &        K1III      &     4527   &    2.48    &   -0.19    & $-0.26^{+0.22}_{-0.24}$ &    0.00    &   58 Leo             \\          
 HD 95418    &  11:01:50.39   &  +56:22:56.4   &          A1V      &     9452   &    3.94    &    0.22    &                         &    0.3     &   $\beta$ UMa        \\          
 HD 95735    &  11:03:20.61   &  +35:58:53.3   &          M2V      &     3739   &    4.96    &   -0.20    & $10.47^{+0.03}_{-0.03}$ &    0.00    &   -                  \\          
 HD 97633    &  11:14:14.44   &  +15:25:47.1   &          A2V      &     9360   &    3.60    &    0.03    & $-0.38^{+0.11}_{-0.11}$ &    0.00    &   $\theta$ Leo       \\          
 HD 97916    &  11:15:54.11   &  +02:05:12.2   &          F5V      &     6308   &    4.08    &   -1.11    & $ 3.61^{+0.36}_{-0.41}$ &    0.00    &   -                  \\
 HD 98262    &  11:18:28.76   &  +33:05:39.3   &        K3III      &     4119   &    1.74    &   -0.11    & $-2.07^{+0.23}_{-0.25}$ &    0.00    &   $\nu$ UMa          \\
 HD 98839    &  11:22:49.61   &  +43:28:57.9   &         G8II      &     4886   &    2.30    &    0.03    & $-0.90^{+0.21}_{-0.23}$ &    0.00    &   56 UMa             \\
 HD 99028    &  11:23:55.37   &  +10:31:46.9   &         F2IV      &     6703   &    3.98    &    0.07    & $ 1.99^{+0.08}_{-0.08}$ &    0.1     &   $\iota$ Leo        \\
 HD 99747    &  11:29:04.70   &  +61:46:40.0   &     F5Vawvar      &     6580   &    4.23    &   -0.57    & $ 3.05^{+0.06}_{-0.06}$ &    0.2     &   -                  \\
 HD 100006   &  11:30:29.08   &  +18:24:35.1   &        K0III      &     4755   &    3.00    &    0.02    & $ 0.56^{+0.20}_{-0.22}$ &    0.00    &   86 Leo             \\
 \hline\noalign{\smallskip}
 \end{tabular}
 \end{flushleft}
\caption[]{ \label{cat3} catalogue of the stars (cont'd)}
 \end{table*}
 \begin{table*}
 \begin{flushleft}
 \begin{tabular}{llllrrrrrl}
 \hline\noalign{\smallskip}
 Star  &   RA 2000   &   DEC 2000   &
  spectr. &  Teff  & log(g)&  [Fe/H]&  Mv    &  Av &     identification  \\
 \noalign{\smallskip}
 \hline\noalign{\smallskip}
 HD 100889   &  11:36:40.95   &  -09:48:08.1   &       B9.5Vn      &            &            &            & $-0.20^{+0.17}_{-0.18}$ &    0.01    &   $\theta$ Crt HR4468  \\
 HD 101501   &  11:41:03.03   &  +34:12:09.2   &       G8Vvar      &     5380   &    4.55    &   -0.11    & $ 5.42^{+0.03}_{-0.03}$ &    0.00    &   61 UMa             \\          
 HD 101606   &  11:41:34.50   &  +31:44:45.5   &          F4V      &     6210   &    4.32    &   -0.79    & $ 2.49^{+0.08}_{-0.08}$ &    0.2     &   62 UMa             \\          
 LTT4364     &  11:45:43      &  -64:50:29     &          DC:      &            &            &            &                         &    0.00    &   -                  \\          
 HD 102212   &  11:45:51.57   &  +06:31:47.3   &        M0III      &     3660   &            &            & $-0.86^{+0.17}_{-0.18}$ &    0.00    &   $\nu$ Vir          \\          
 HD 102224   &  11:46:03.13   &  +47:46:45.6   &        K0III      &     4358   &    1.61    &   -0.46    & $-0.20^{+0.09}_{-0.10}$ &    0.00    &   $\chi$ UMa         \\          
 HD 102634   &  11:49:01.40   &  -00:19:07.2   &          F7V      &     6314   &    4.12    &    0.18    & $ 3.47^{+0.09}_{-0.09}$ &    0.00    &   -                  \\          
 HD 102870   &  11:50:41.29   &  +01:45:55.4   &          F8V      &     6121   &    4.13    &    0.15    & $ 3.31^{+0.04}_{-0.04}$ &    0.10    &   $\beta$ Vir        \\          
 HD 103036   &  11:51:50.13   &  -05:45:43.8   &     G3Ibpvar      &     4275   &    0.49    &   -1.64    &                         &    0.5     &   V* TY Vir          \\          
 HD 104893   &  12:04:43.15   &  -29:11:05.1   &          K0I      &     4500   &    0.96    &   -1.90    &                         &    0.00    &   -                  \\          
 HD 105546   &  12:09:02.76   &  +59:01:05.7   &       G2IIIw      &     5224   &    2.39    &   -1.37    &                         &    0.6     &   -                  \\          
 HD 106038   &  12:12:01.50   &  +13:15:44.5   &       F6V-VI      &     5988   &    4.41    &   -1.35    & $ 4.55^{+0.38}_{-0.44}$ &    0.4     &   -                  \\          
 HD 107213   &  12:19:29.66   &  +28:09:26.0   &         F8Vs      &     6318   &    4.06    &    0.26    & $ 2.90^{+0.10}_{-0.10}$ &    0.00    &   9 Com              \\          
 HD 108177   &  12:25:34.98   &  +01:17:06.4   &         sdF5      &     6069   &    4.33    &   -1.69    & $ 4.82^{+0.28}_{-0.32}$ &    0.06    &   -                  \\          
 HD 109995   &  12:38:47.69   &  +39:18:32.9   &          A0p      &     8293   &    3.16    &   -1.27    & $ 1.04^{+0.38}_{-0.45}$ &    0.0     &   -                  \\          
 HD 110897   &  12:44:59.68   &  +39:16:42.9   &          G0V      &     5845   &    4.24    &   -0.48    & $ 4.66^{+0.04}_{-0.04}$ &    0.1     &   10 CVn             \\          
 HD 111028   &  12:46:22.38   &  +09:32:26.8   &     K1III-IV      &     4710   &    3.00    &   -0.40    & $ 2.41^{+0.11}_{-0.12}$ &    0.00    &   33 Vir             \\          
 HD 112300   &  12:55:36.48   &  +03:23:51.4   &        M3III      &     3684   &    1.30    &   -0.14    & $-0.55^{+0.13}_{-0.14}$ &    0.00    &   $\delta$ Vir       \\          
 HD 113022   &  13:00:38.86   &  +18:22:22.3   &         F6Vs      &     6380   &    4.20    &    0.10    & $ 3.15^{+0.09}_{-0.09}$ &    0.00    &   -                  \\          
 HD 113139   &  13:00:43.59   &  +56:21:58.8   &          F2V      &            &            &            & $ 2.40^{+0.05}_{-0.05}$ &    0.6     &   78 UMa             \\          
 HD 113337   &  13:01:47.15   &  +63:36:36.6   &          F6V      &     6545   &    4.20    &    0.07    & $ 3.05^{+0.06}_{-0.06}$ &    0.1     &   -                  \\          
 HD 113226   &  13:02:10.76   &  +10:57:32.8   &     G8IIIvar      &     5017   &    2.79    &    0.09    & $ 0.34^{+0.08}_{-0.08}$ &    0.00    &   $\varepsilon$ Vir  \\          
 HD 113848   &  13:06:21.28   &  +21:09:12.6   &          F4V      &     6628   &    4.03    &   -0.27    & $ 2.47^{+0.16}_{-0.17}$ &    0.1     &   39 Com             \\          
 HD 114330   &  13:09:57.01   &  -05:32:20.1   &          A1V      &     9509   &    3.67    &   -0.02    & $-1.13^{+0.30}_{-0.35}$ &    0.00    &   $\theta$ Vir HR4963  \\        
 HD 115383   &  13:16:46.71   &  +09:25:25.3   &         G0Vs      &     5981   &    4.10    &    0.08    & $ 3.73^{+0.05}_{-0.05}$ &    0.19    &   59 Vir             \\          
 HD 116842   &  13:25:13.42   &  +54:59:16.8   &          A5V      &     8066   &            &            & $ 1.61^{+0.05}_{-0.05}$ &    0.4     &   80 UMa ALCOR       \\          
 HD 117176   &  13:28:25.95   &  +13:46:48.7   &          G5V      &     5485   &    3.84    &   -0.09    & $ 3.68^{+0.05}_{-0.05}$ &    0.00    &   70 Vir             \\          
 HD 120136   &  13:47:16.04   &  +17:27:24.4   &          F7V      &     6445   &    4.22    &    0.27    & $ 3.52^{+0.04}_{-0.04}$ &    0.00    &   $\tau$ Boo         \\          
 HD 122408   &  14:01:38.78   &  +01:32:40.5   &          A3V      &     8290   &            &            & $ 0.10^{+0.14}_{-0.15}$ &    0.01    &   $\tau$ Vir         \\          
 HD 122563   &  14:02:31.96   &  +09:41:10.6   &         F8IV      &     4588   &    1.21    &   -2.66    & $-0.91^{+0.40}_{-0.48}$ &    0.00    &   -                  \\          
 HD 123299   &  14:04:23.43   &  +64:22:32.9   &        A0III      &     9927   &    3.17    &   -0.33    & $-1.25^{+0.12}_{-0.13}$ &    0.00    &   $\alpha$ Dra       \\          
 HD 124425   &  14:13:40.67   &  -00:50:42.4   &         F7Vw      &     6380   &    4.04    &    0.09    & $ 2.12^{+0.11}_{-0.12}$ &    0.05    &   -                  \\          
 HD 124570   &  14:14:05.33   &  +12:57:34.5   &         F6IV      &     6192   &    4.06    &    0.07    & $ 2.93^{+0.07}_{-0.07}$ &    0.00    &   14 Boo             \\          
 HD 124850   &  14:16:00.88   &  -05:59:58.3   &          F7V      &     6143   &    3.89    &   -0.13    & $ 2.34^{+0.06}_{-0.06}$ &    0.08    &   $\iota$ Vir        \\
 HD 125560   &  14:19:45.32   &  +16:18:24.5   &        K3III      &     4426   &    2.42    &    0.00    & $ 0.72^{+0.10}_{-0.11}$ &    0.3     &   20 Boo             \\
 HD 126141   &  14:23:06.92   &  +25:20:16.9   &          F5V      &     6632   &    4.30    &    0.00    & $ 3.45^{+0.08}_{-0.08}$ &    0.00    &   -                  \\
 HD 126660   &  14:25:12.02   &  +51:51:06.2   &          F7V      &     6369   &    4.29    &   -0.05    & $ 3.22^{+0.03}_{-0.03}$ &    0.01    &   $\theta$ Boo       \\
 HD 127665   &  14:31:49.86   &  +30:22:16.1   &        K3III      &     4251   &    1.95    &   -0.25    & $-0.43^{+0.10}_{-0.10}$ &    0.7     &   $\rho$ Boo         \\
 HD 128167   &  14:34:40.69   &  +29:44:41.3   &      F3Vwva       &     6746   &    4.23    &   -0.42    & $ 3.17^{+0.04}_{-0.04}$ &    0.35    &   $\sigma$ Boo       \\
 HD 130109   &  14:46:14.99   &  +01:53:34.6   &          A0V      &    10110   &            &            & $ 0.72^{+0.09}_{-0.09}$ &    0.02    &   109 Vir HR5511     \\
 HD 130948   &  14:50:15.72   &  +23:54:42.4   &          G2V      &     5947   &    4.07    &    0.07    & $ 4.61^{+0.05}_{-0.05}$ &    0.00    &   -                  \\
 HD 131156   &  14:51:23.28   &  +19:06:02.3   &          G8V      &     5512   &    4.59    &   -0.04    & $ 5.55^{+0.03}_{-0.03}$ &    0.00    &   $\xi$ Boo          \\
 \hline\noalign{\smallskip}
 \end{tabular}
 \end{flushleft}
\caption[]{ \label{cat4} catalogue of the stars (cont'd)}
 \end{table*}
 \begin{table*}
 \begin{flushleft}
 \begin{tabular}{llllrrrrrl}
 \hline\noalign{\smallskip}
 Star  &   RA 2000   &   DEC 2000   &
  spectr. &   Teff  &  log(g)&  [Fe/H]& Mv    & Av &      identification  \\
 \noalign{\smallskip}
 \hline\noalign{\smallskip}
 HD 132142   &  14:55:12.00   &  +53:40:45.1   &          K1V      &     5125   &    4.50    &   -0.55    & $ 5.87^{+0.05}_{-0.05}$ &    0.00    &   -                  \\
 HD 134083   &  15:07:17.95   &  +24:52:10.5   &          F5V      &     6583   &    4.20    &    0.06    & $ 3.45^{+0.05}_{-0.05}$ &    0.01    &   45 Boo             \\          
 HD 134169   &  15:08:18.06   &  +03:55:50.3   &         G1Vw      &     5827   &    3.97    &   -0.87    & $ 3.82^{+0.17}_{-0.18}$ &    0.00    &   -                  \\          
 HD 135722   &  15:15:30.10   &  +33:18:54.4   &        G8III      &     4803   &    2.51    &   -0.40    & $ 0.69^{+0.06}_{-0.06}$ &    0.00    &   $\delta$ Boo       \\          
 HD 136202   &  15:19:18.58   &  +01:46:00.0   &     F8III-IV      &     6068   &    3.94    &   -0.12    & $ 3.06^{+0.06}_{-0.06}$ &    0.04    &   5 Ser V* MQ Ser    \\          
 HD 136512   &  15:20:08.64   &  +29:36:58.7   &        K0III      &     4686   &    2.72    &   -0.40    & $ 0.89^{+0.15}_{-0.16}$ &    0.00    &   o CrB              \\          
 HD 137759   &  15:24:55.78   &  +58:57:57.7   &        K2III      &     4472   &    2.74    &    0.19    & $ 0.82^{+0.05}_{-0.05}$ &    0.00    &   $\iota$ Dra        \\          
 HD 138290   &  15:30:55.42   &  +08:34:44.8   &         F4Vw      &     6834   &    4.11    &   -0.05    & $ 3.05^{+0.12}_{-0.12}$ &    0.00    &   -                  \\          
 HD 139669   &  15:31:25.05   &  +77:20:57.6   &        K5III      &     3907   &    1.69    &   -0.11    & $-2.03^{+0.28}_{-0.31}$ &    0.00    &   $\theta$ UMi       \\          
 HD 139641   &  15:37:49.55   &  +40:21:11.8   &     G8III-IV      &     4937   &    3.21    &   -0.54    & $ 1.77^{+0.08}_{-0.08}$ &    0.00    &   $\varphi$ Boo      \\          
 HD 139798   &  15:38:16.14   &  +46:47:52.8   &          F2V      &     6788   &    4.00    &   -0.13    & $ 2.98^{+0.06}_{-0.06}$ &    0.00    &   -                  \\          
 HD 141004   &  15:46:26.75   &  +07:21:11.7   &       G0Vvar      &     5899   &    4.22    &   -0.01    & $ 4.07^{+0.04}_{-0.04}$ &    0.00    &   $\lambda$ Ser      \\          
 HD 141714   &  15:49:35.70   &  +26:04:06.8   &     G5III-IV      &     5218   &    3.12    &   -0.31    & $ 1.08^{+0.10}_{-0.10}$ &    0.00    &   $\delta$ CrB       \\          
 HD 142373   &  15:52:40.19   &  +42:27:00.0   &          F9V      &     5823   &    4.11    &   -0.41    & $ 3.60^{+0.04}_{-0.04}$ &    0.01    &   $\chi$ Her         \\          
 HD 144206   &  16:02:47.85   &  +46:02:12.7   &        B9III      &    11957   &    3.70    &   -0.17    & $-0.95^{+0.16}_{-0.16}$ &    0.35    &   $\upsilon$ Her     \\          
 HD 145675   &  16:10:24.21   &  +43:49:06.1   &          K0V      &     5312   &    4.40    &    0.46    & $ 5.32^{+0.04}_{-0.04}$ &    0.00    &   14 Her             \\          
 HD 145976   &  16:12:45.43   &  +26:40:14.4   &          F3V      &     6720   &    4.10    &    0.01    & $ 2.18^{+0.14}_{-0.15}$ &    0.00    &   -                  \\          
 HD 146051   &  16:14:20.77   &  -03:41:38.3   &        M1III      &     3679   &    1.40    &    0.32    & $-0.86^{+0.13}_{-0.14}$ &    0.00    &   $\delta$ Oph       \\          
 HD 147394   &  16:19:44.45   &  +46:18:47.8   &         B5IV      &    14908   &    3.81    &   -0.07    & $-1.13^{+0.13}_{-0.13}$ &    0.08    &   $\tau$ Her         \\          
 HD 147547   &  16:21:55.24   &  +19:09:10.9   &        A9III      &     7125   &            &            & $-0.23^{+0.11}_{-0.11}$ &    0.08    &   V* $\gamma$ Her    \\          
 HD 148513   &  16:28:33.98   &  +00:39:54.6   &       K4IIIp      &     4003   &    1.03    &   -0.08    & $-0.16^{+0.25}_{-0.28}$ &    0.00    &   -                  \\          
 HD 148783   &  16:28:38.52   &  +41:52:54.1   &     M6IIIvar      &     3250   &    0.20    &   -0.02    & $-1.31^{+0.16}_{-0.17}$ &    1.0     &   V* g Her 30 Her    \\          
 HD 150275   &  16:30:39.08   &  +77:26:45.1   &        K1III      &     4667   &    2.50    &   -0.50    & $ 0.87^{+0.17}_{-0.18}$ &    0.00    &   -                  \\          
 HD 149121   &  16:32:35.68   &  +05:31:16.4   &      B9.5III      &    11067   &    3.78    &    0.17    & $ 0.64^{+0.21}_{-0.23}$ &    0.00    &   28 Her             \\          
 HD 149661   &  16:36:21.18   &  -02:19:25.8   &          K2V      &     5362   &    4.56    &    0.01    & $ 5.82^{+0.04}_{-0.04}$ &    0.00    &   12 Oph V* V2133 Oph\\          
 HD 151044   &  16:42:27.69   &  +49:56:12.1   &          F8V      &     6110   &    4.40    &   -0.02    & $ 4.14^{+0.05}_{-0.05}$ &    0.00    &   -                  \\          
 HD 151217   &  16:45:49.89   &  +08:34:57.3   &        K5III      &     4093   &    3.36    &   -0.04    & $ 0.00^{+0.21}_{-0.22}$ &    0.00    &   43 Her             \\          
 HD 152830   &  16:55:15.97   &  +13:37:12.0   &         F5II      &     6811   &    3.70    &   -0.13    & $ 1.87^{+0.15}_{-0.15}$ &    0.00    &   V* V644 Her        \\          
 HD 154417   &  17:05:16.83   &  +00:42:12.1   &          F9V      &     5859   &    4.27    &   -0.11    & $ 4.45^{+0.06}_{-0.06}$ &    0.00    &   V* V2213 Oph       \\          
 HD 154733   &  17:06:18.11   &  +22:05:03.3   &        K4III      &     4220   &    2.20    &   -0.13    & $ 0.37^{+0.19}_{-0.20}$ &    0.00    &   -                  \\          
 HD 155646   &  17:12:54.33   &  +00:21:07.9   &        F6III      &     6168   &    3.94    &   -0.14    & $ 2.47^{+0.16}_{-0.17}$ &    0.00    &   -                  \\
 HD 156283   &  17:15:02.85   &  +36:48:33.0   &      K3IIvar      &     4128   &    1.68    &   -0.18    & $-2.12^{+0.14}_{-0.15}$ &    0.00    &   $\pi$ Her          \\
 HD 157373   &  17:20:33.60   &  +48:11:19.7   &     F6Vawvar      &     6433   &    4.09    &   -0.50    & $ 3.36^{+0.08}_{-0.08}$ &    0.00    &   -                  \\
 HD 157214   &  17:20:39.47   &  +32:28:13.0   &          G0V      &     5686   &    4.26    &   -0.36    & $ 4.60^{+0.03}_{-0.03}$ &    0.00    &   72 Her             \\
 HD 157089   &  17:21:07.15   &  +01:26:32.6   &          F9V      &     5786   &    4.13    &   -0.55    & $ 4.03^{+0.10}_{-0.10}$ &    0.00    &   -                  \\
 HD 157856   &  17:25:57.84   &  -01:39:06.9   &          F3V      &     6327   &    3.96    &   -0.18    & $ 2.50^{+0.12}_{-0.13}$ &    0.00    &   -                  \\
 HD 157881   &  17:25:45.57   &  +02:06:51.5   &          K7V      &     4073   &    4.58    &    0.16    & $ 8.12^{+0.04}_{-0.04}$ &    0.00    &   -                  \\
 HD 157999   &  17:26:30.88   &  +04:08:25.2   &      K3IIvar      &     4146   &            &    0.01    &                         &    0.00    &   $\sigma$ Oph       \\
 HD 159181   &  17:30:25.98   &  +52:18:04.9   &         G2II      &     5169   &    1.55    &    0.14    & $-3.85^{+0.13}_{-0.14}$ &    1.4     &   $\beta$ Dra        \\
 HD 159332   &  17:33:22.84   &  +19:15:24.8   &          F6V      &     6172   &    3.88    &   -0.23    & $ 2.83^{+0.07}_{-0.08}$ &    0.00    &   -                  \\
 HD 160693   &  17:39:37.24   &  +37:11:08.7   &          G0V      &     5747   &    4.21    &   -0.63    & $ 4.71^{+0.11}_{-0.12}$ &    0.00    &   -                  \\
 HD 160910   &  17:41:58.64   &  +15:57:07.8   &         F4Vw      &     6502   &    4.00    &   -0.13    & $ 2.81^{+0.08}_{-0.08}$ &    0.00    &   -                  \\
 \hline\noalign{\smallskip}
 \end{tabular}
 \end{flushleft}
\caption[]{ \label{cat5} catalogue of the stars (cont'd)}
 \end{table*}
 \begin{table*}
 \begin{flushleft}
 \begin{tabular}{llllrrrrrl}
 \hline\noalign{\smallskip}
 Star  &   RA 2000   &   DEC 2000   &
  spectr. &   Teff  &  log(g)&  [Fe/H]& Mv    & Av &      identification  \\
 \noalign{\smallskip}
 \hline\noalign{\smallskip} 
 HD 161817   &  17:46:40.65   &  +25:44:57.3   &         sdA2      &     7702   &    2.94    &   -1.30    & $ 0.80^{+0.25}_{-0.28}$ &    0.0     &   -                  \\
 HD 163506   &  17:55:25.19   &  +26:02:59.9   &      F2Iavar      &     6400   &    1.20    &   -0.41    &                         &    0.27    &   89 Her V* V441 Her \\
 HD 163993   &  17:57:45.83   &  +29:14:52.5   &        K0III      &     5014   &    2.78    &    0.02    & $ 0.61^{+0.06}_{-0.06}$ &    0.00    &   $\xi$ Her          \\          
 HD 164136   &  17:58:30.15   &  +30:11:21.4   &         F2II      &     6586   &    2.07    &   -0.42    & $-2.82^{+0.29}_{-0.33}$ &    0.30    &   $\nu$. Her V*      \\          
 HD 164349   &  18:00:03.42   &  +16:45:03.4   &     K0II-III      &     4383   &    1.80    &   -0.22    & $-1.85^{+0.35}_{-0.41}$ &    0.00    &   93 Her             \\          
 HD 164353   &  18:00:38.72   &  +02:55:53.7   &         B5Ib      &    22150   &    3.73    &   -0.16    &                         &    0.35    &   67 Oph             \\          
 HD 165195   &  18:04:40.09   &  +03:46:45.4   &          K3p      &     4450   &    1.31    &   -2.16    &                         &    0.00    &   -                  \\          
 HD 165341   &  18:05:27.21   &  +02:30:08.8   &          K0V      &     5025   &    4.69    &   -0.20    & $ 5.60^{+0.03}_{-0.03}$ &    0.00    &   70 Oph V* V2391 Oph\\          
 HD 165908   &  18:07:01.61   &  +30:33:42.7   &          F7V      &     5952   &    4.26    &   -0.56    & $ 4.02^{+0.04}_{-0.04}$ &    0.08    &   99 Her             \\          
 HD 166620   &  18:09:37.65   &  +38:27:32.1   &          K2V      &     4955   &    4.50    &   -0.25    & $ 5.86^{+0.03}_{-0.03}$ &    0.3     &   -                  \\          
 HD 166285   &  18:09:54.01   &  +03:07:13.1   &          F5V      &     6271   &    3.90    &   -0.22    & $ 2.32^{+0.09}_{-0.09}$ &    0.00    &   -                  \\          
 HD 168151   &  18:13:53.36   &  +64:23:49.9   &          F5V      &     6463   &    4.05    &   -0.30    & $ 2.64^{+0.04}_{-0.04}$ &    0.5     &   36 Dra             \\          
 HD 167858   &  18:17:04.85   &  +01:00:20.9   &          F2V      &     7041   &    4.00    &    0.17    & $ 2.63^{+0.13}_{-0.13}$ &    0.00    &   -                  \\          
 HD 172167   &  18:36:56.19   &  +38:46:58.8   &       A0Vvar      &     9520   &    3.97    &   -0.55    &                         &    0.01    &   $\alpha$ Lyr Vega HR7001  \\   
 HD 173780   &  18:46:04.47   &  +26:39:43.5   &        K3III      &     4413   &    2.57    &   -0.11    & $ 0.39^{+0.11}_{-0.11}$ &    0.00    &   -                  \\          
 HD 173880   &  18:47:01.22   &  +18:10:52.4   &        A5III      &     8113   &            &   -0.84    & $ 2.07^{+0.06}_{-0.06}$ &    0.0     &   111 Her            \\          
 HD 175305   &  18:47:05.73   &  +74:43:30.8   &        G5III      &     5046   &    2.55    &   -1.44    & $ 1.15^{+0.21}_{-0.23}$ &    0.00    &   -                  \\          
 HD 173819   &  18:47:28.98   &  -05:42:18.3   &     K0Ibpvar      &     4421   &    0.00    &   -0.88    &                         &    0.00    &   V* R Sct           \\          
 HD 175640   &  18:56:22.66   &  -01:47:59.3   &        B9III      &    12067   &    3.93    &   -0.47    & $-0.04^{+0.26}_{-0.29}$ &    0.2     &   -                  \\          
 HD 176437   &  18:58:56.62   &  +32:41:22.4   &        B9III      &    10080   &    3.50    &    0.11    & $-3.73^{+0.22}_{-0.24}$ &    0.5     &   $\gamma$ Lyr       \\          
 HD 176303   &  18:59:05.73   &  +13:37:21.2   &          F8V      &     6122   &    4.22    &   -0.06    & $ 1.88^{+0.09}_{-0.09}$ &    0.00    &   11 Aql             \\          
 HD 181470   &  19:19:01.15   &  +37:26:43.1   &        A0III      &     7887   &    3.53    &   -0.32    & $-0.69^{+0.25}_{-0.27}$ &    0.4     &   -                  \\          
 HD 182101   &  19:22:48.35   &  +09:54:46.4   &          F6V      &     6303   &    4.18    &   -0.26    & $ 3.41^{+0.09}_{-0.09}$ &    0.18    &   -                  \\          
 HD 182490   &  19:24:22.08   &  +16:56:15.9   &     A2III-IV      &            &            &            & $ 0.86^{+0.23}_{-0.26}$ &    0.2     &   2 Sge              \\          
 HD 338529   &  19:32:31.91   &  +26:23:27.6   &           B5      &     6192   &    3.79    &   -2.29    & $ 3.58^{+0.45}_{-0.55}$ &    0.00    &   -                  \\          
 HD 184960   &  19:34:19.76   &  +51:14:13.5   &          F7V      &     6249   &    4.36    &   -0.14    & $ 3.38^{+0.04}_{-0.04}$ &    0.3     &   -                  \\          
 HD 184927   &  19:35:32.01   &  +31:16:35.9   &          B2V      &    21913   &    3.80    &    0.41    &                         &    0.25    &   V* V1671 Cyg       \\          
 HD 185395   &  19:36:26.54   &  +50:13:13.7   &          F4V      &     6704   &    4.35    &   -0.02    & $ 3.15^{+0.04}_{-0.04}$ &    0.00    &   $\theta$ Cyg       \\          
 HD 185657   &  19:37:56.68   &  +49:17:02.6   &          G6V      &     4912   &    3.00    &   -0.41    & $ 0.22^{+0.20}_{-0.22}$ &    0.30    &   -                  \\          
 HD 188209   &  19:51:59.07   &  +47:01:38.5   &       O9.5Ia      &    32903   &    3.32    &            &                         &    0.70    &   HR7589             \\          
 HD 187929   &  19:52:28.36   &  +01:00:20.4   &        F6Ib:      &     5826   &    1.50    &    0.07    &                         &    0.00    &   $\eta$ Aql V*      \\          
 HD 188665   &  19:53:17.37   &  +57:31:24.5   &          B5V      &            &            &            & $-1.49^{+0.22}_{-0.24}$ &    0.16    &   23 Cyg             \\          
 HD 188510   &  19:55:09.70   &  +10:44:24.9   &        G5Vwe      &     5536   &    4.59    &   -1.63    & $ 5.85^{+0.13}_{-0.13}$ &    0.00    &   2.52               \\          
 HD 189849   &  20:01:06.01   &  +27:45:12.8   &        A4III      &     7860   &    3.61    &    0.01    & $ 0.49^{+0.11}_{-0.11}$ &    0.00    &   15 Vul V* NT Vul   \\          
 HD 194093   &  20:22:13.70   &  +40:15:24.1   &         F8Ib      &     5253   &    0.99    &   -0.07    & $-6.42^{+0.48}_{-0.61}$ &    0.28    &   $\gamma$ Cyg       \\          
 HD 195725   &  20:29:34.83   &  +62:59:38.9   &        A7III      &     7769   &    4.00    &    0.13    & $ 0.31^{+0.06}_{-0.07}$ &    0.8     &   $\theta$ Cep       \\          
 HD 195986   &  20:32:52.33   &  +43:11:29.7   &        B4III      &            &            &            & $-2.36^{+0.57}_{-0.76}$ &    0.25    &   -                  \\          
 HD 197345   &  20:41:25.91   &  +45:16:49.2   &         A2Ia      &     8582   &    1.13    &    0.48    &                         &    0.36    &   $\alpha$ Cyg Deneb HD 197345 \\ 
 Feige110    &  23:19:58.40   &  -05:09:56.2   &          DA:      &            &            &            &                         &    0.00    &   -                  \\
 \hline\noalign{\smallskip}
\end{tabular}
 \end{flushleft}
\caption[]{ \label{cat6} catalogue of the stars (cont'd)}
 \end{table*}

\section{The library}{\label{sect5}}
Once calibrated, the spectra in the 4 (JKT) or 6 (SSO) settings of the program
stars were combined by averaging the overlapping pixels. This results in 257
stellar spectra in {\it fits} format resampled with a step of 1\AA\ per pixel.
The library is available in two different forms: the "raw" data, including the
combined spectra as coming out of the calibration process, and the data corrected 
for interstellar reddening using the empirical extinction function of Cardelli et 
al. (\cite{Cardelli}). As an additional check, we have compared the reddening 
corrected spectra (using the extintion values from the literature) to the 
equivalent ones in the Kurucz atlas (same spectral type and metallicity). In most 
cases ($\sim$80\%), the agreement between the two spectra is excellent. In case 
of discrepancy ($\sim$20\% of the sample), the extintion values used and quoted 
in the tables are those allowing to match our corrected spectra to Kurucz. These
discrepant objects are clearly identifyed on the web site.

\section{Synthetic spectra of stellar populations}{\label{sect6}}
Templates of stellar populations have been built from the dereddened library. At 
this stage, the wavelength scale was corrected for radial velocity. For some 
stars, data are missing in limited wavelength ranges: for these, we filled the 
gaps using spectra of stars of similar or close spectral type. The final spectra 
are useful from 3200 to 9300\AA\ because spectra become noisy from 9300\AA\ to 
9850\AA. In this way, we built several subsets of the atlas with different [Fe/H] 
ranges (Table~\ref{tab5}). These subsets include a total of 242 star templates 
which are also available on the web site.

\begin{table}
\centering
\caption[]{\label{tab5} Subsets of stars according to their metallicity}
\begin{flushleft}
\begin{tabular}{llr}
\hline\noalign{\smallskip}
   &  [Fe/H]  range &  number of stars \\
\noalign{\smallskip}
\hline\noalign{\smallskip}
           &    $<$ -1.90       &    6  \\
           &  [-1.90 , -1.41]   &   12  \\
sub-solar  &  [-1.40 , -0.96]   &    6  \\
           &  [-0.95 , -0.49]   &   23  \\
           &  [-0.48 , -0.06]   &   69  \\
\hline\noalign{\smallskip}
solar      &  [-0.06 , +0.07]   &   84  \\
\hline\noalign{\smallskip}
 above     &  [+0.08 , +0.47]   &   38  \\
 solar     &    $>$ +0.47       &    4  \\
\noalign{\smallskip}
\hline
\end{tabular}
\end{flushleft}
\end{table}

\subsection{Comparison with Kurucz spectra}
\begin{figure}
\includegraphics[width=7cm]{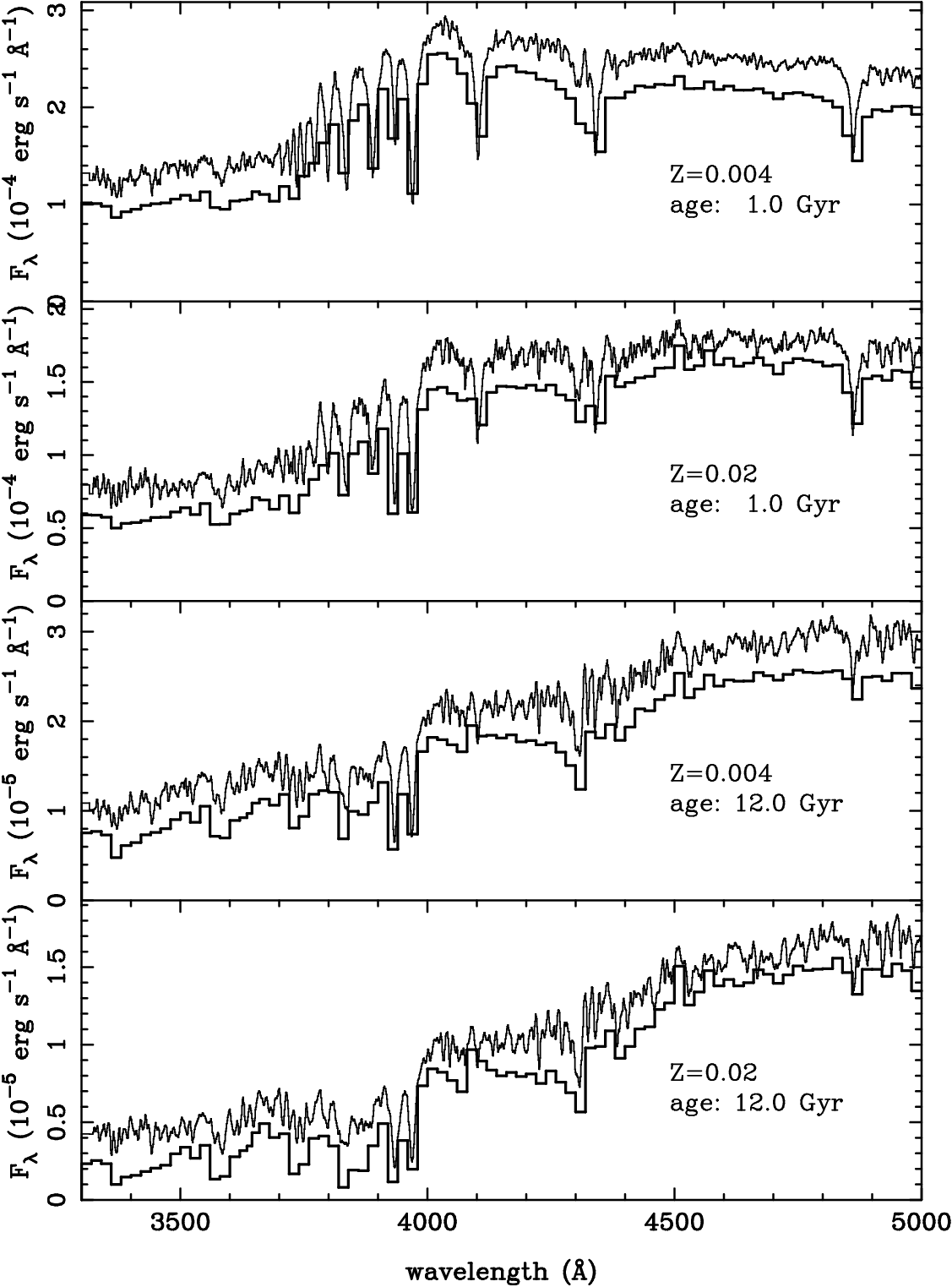}  %  {sed_4000.ps}}
\caption{Synthetic spectra built with STELIB compared to synthetic spectra
obtained with Kurucz spectra using GISSEL02. SSP, wavelength range 3300\AA\
to 5000\AA, Z=0.02 (solar metallicity) and  Z=0.004,
from 1 Gyr to 12 Gyr. The Kurucz spectra are shifted downward by 10\% of the scale for
clarity.  } \label{fig8}
\end{figure}

\begin{figure}
\includegraphics[width=7cm]{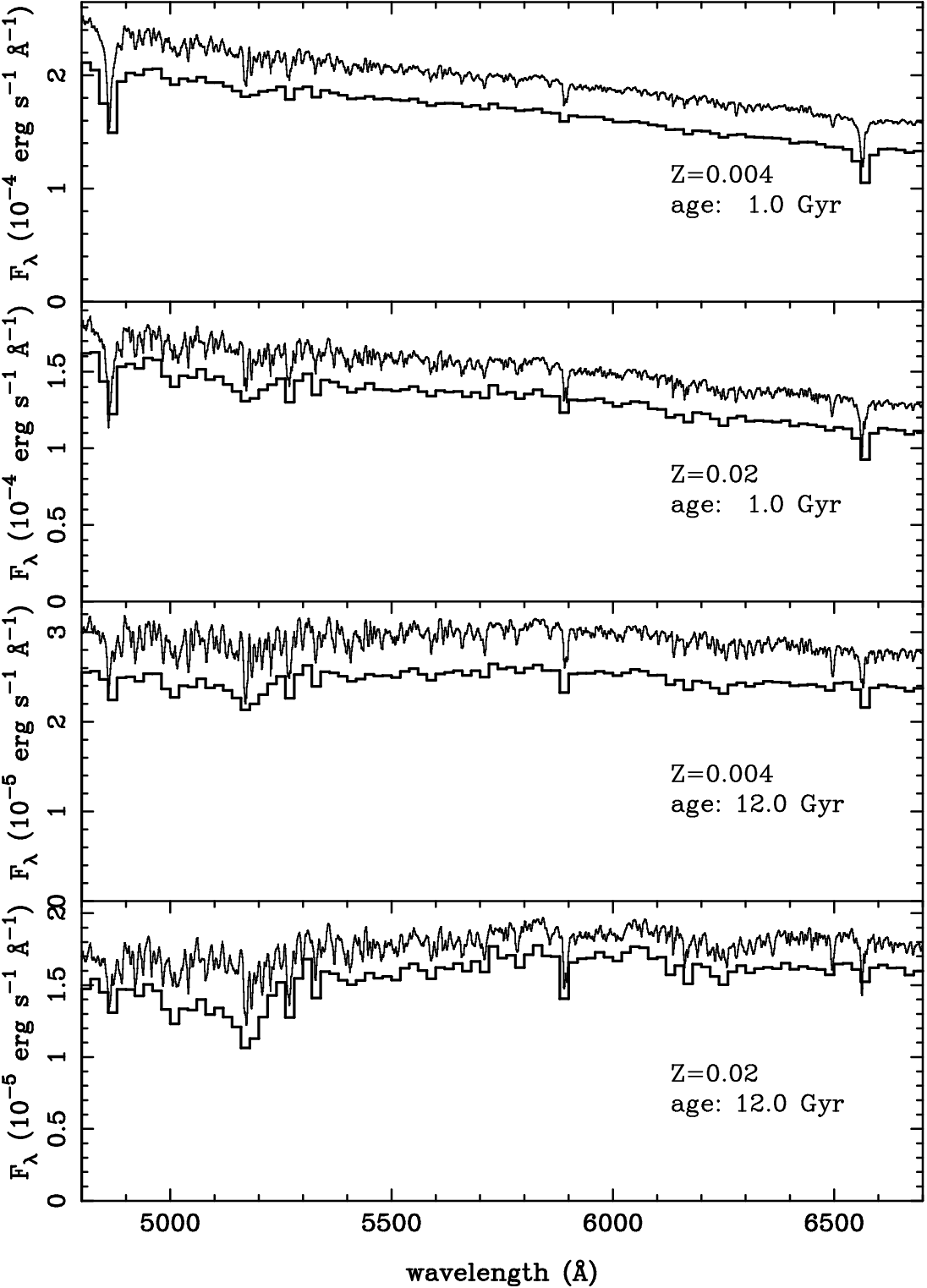}  %  {sed_6000.ps}}
\caption{Synthetic spectra built with STELIB compared to synthetic spectra
obtained with Kurucz spectra using GISSEL02. SSP, wavelength range 4800\AA\
to 7000\AA, Z=0.02 (solar metallicity) and  Z=0.004,
from 1 Gyr to 12 Gyr. The Kurucz spectra are shifted downward by 10\% of the scale for
clarity.   } \label{fig9}
\end{figure}

In order to compare with the previous results, and to show the capabilities of 
this new library, we have built galaxy models with the STELIB library in the
apropriate subset, using the new galaxy evolutionary code GISSEL02 (Bruzual
\& Charlot \cite{Bruzual}). The evolution of a single stellar population of solar 
metallicity (Z=0.02) is given in Fig.~\ref{fig8} and \ref{fig9}, for ages of the 
stellar population ranging from 1 to 12 Gyr. A comparison with the same models 
obtained with the synthetic stellar spectra from Kurucz is also shown to 
emphasize the gain in spectral resolution.

\subsection{Modeling observed spectra of galaxies}

An important application of STELIB is to reproduce in details the spectral 
features observed in galaxies at z\ltapprox 1. As an example, we present here 
the modeling of spectra of galaxies belonging to the cluster AC114 (more 
officially named \object{ACO S 1077}, Abell et al., \cite{Abell}) at a redshift 
of z=0.312, and some foreground galaxies in the same field. These spectra were 
obtained with the spectrograph FORS1 on VLT unit 1 Antu, on october 5, 1999. The 
main objective of the run was the determination of the redshift of background 
lensed galaxies, but spectra of cluster galaxies were also obtained in the 
remaining slits. The grism used was G300V, with a wavelength coverage between 
$\sim$4000\AA\ and $\sim$8600\AA, and a wavelength resolution of R=500 for the 1" 
slit width used which correspond to a resolution of $\sim$7\AA\ at rest frame of 
the cluster. Details of the observation conditions and data reduction can be 
found in Campusano et al. (\cite{Campusano}).

\begin{table*}
\centering
\caption[]{\label{tab6} list of modeled galaxy spectra in the field of AC114}
\begin{flushleft}
\begin{tabular}{lclllllr}
\hline\noalign{\smallskip}
     USNO  &  ident.in   &   ra(2000)   &   dec(2000)   &  exposure  & redshift & spectral & age of \\
           &fig. \ref{fig10}&           &               &   time     &          &   type   & model  \\
\noalign{\smallskip}
\hline\noalign{\smallskip}
  U0525\_44063178 &  a &  22:59:00.61 &  -34:46:20.8  &   2700s    &  0.1418 &elliptical&   12 Gyr \\
                  &  b &  22:58:37.19 &  -34:49:27.8  &   5400s    &  0.2605 & irregular&  500 Myr \\
  U0525\_44063148 &  c &  22:58:43.07 &  -34:48:48.1  &   5400s    &  0.2999 & irregular&  500 Myr \\
                  &  d &  22:58:56.93 &  -34:47:57.4  &   2700s    &  0.3113 &  spiral  &    1 Gyr \\
                  &  e &  22:58:46.51 &  -34:49:09.6  &   4218s    &  0.3139 &elliptical&   12 Gyr \\
  U0525\_44062505 &  f &  22:58:37.67 &  -34:49:24.8  &   4218s    &  0.3147 &elliptical&   12 Gyr \\
  U0525\_44063303 &  g &  22:58:44.48 &  -34:49:11.3  &   4218s    &  0.3163 &elliptical&   12 Gyr \\
  U0525\_44063178 &  h &  22:58:43.35 &  -34:49:36.5  &   5400s    &  0.3207 & irregular&  500 Myr \\
\noalign{\smallskip}
\hline
\end{tabular}
\end{flushleft}
\end{table*}

\begin{figure*} [h]
 \centerline{
\includegraphics[width=16.cm,angle=0]{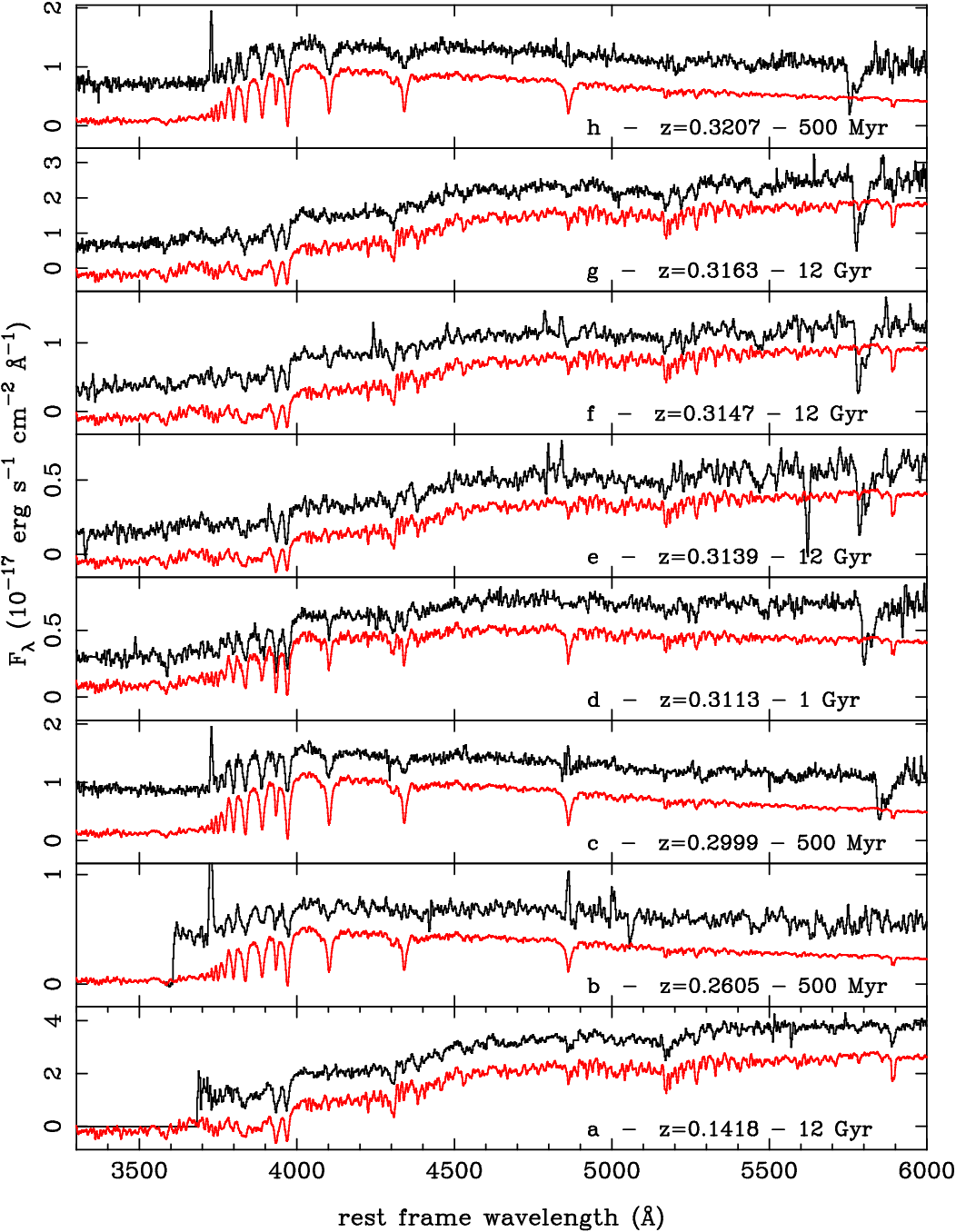}  
 }
\caption{Comparison of galaxy spectra, in and foreground of the cluster of 
galaxies AC114, with solar metallicity SSP synthetic spectra. See 
Table~\ref{tab6} for identification.  The observed spectra are not corrected for 
atmospheric molecular bands (the main one appears at about 5800\AA\ for 
z$\sim$0.3 galaxies)). The observed spectra appear as a black line, and the model 
as a grey line. The model spectra are shifted downward by 20\% of the scale for 
clarity. The legends in the figure give the observed redshift of the galaxies and 
the age of the models. The contribution of emission lines in the Balmer series 
is visible on observed spectra b, c and h.}
\label{fig10}
\end{figure*}

A simple best fit procedure has been used to determine the spectral type of each 
galaxy, acording to its spectral features (see table \ref{tab6}). We have chosen 
to display galaxies of different types, from E to irregulars, with good S/N 
ratio. The models correspond to the evolution with time of a Single Stellar 
Population (SSP) built with the STELIB library for solar metallicity, assuming 
Kroupa (\cite{Kroupa}) IMF. The comparison between observed and modeled spectra 
is shown in Fig.~\ref{fig10}. The best model was chosen among SEDs computed at 11 
different ages from 100  Myr to 12 Gyr. The gain in spectral resolution is clear, 
with obvious applications in stellar population synthesis modelling. In 
particular, STELIB allows to determine the stellar populations using the 
strengths of a large number of absorption lines, due to the wide spectral 
coverage, and thus to improve the emission-line measurements in star-forming 
galaxies and AGNs.

\section{Conclusions}{\label{sect7}}
We have presented the main characteristics of the public stellar library STELIB, 
available on the web site {\it http://webast.ast.obs-mip.fr/stelib}. The main 
improvements with respect to other previous libraries are:
\begin{enumerate}
\item The homogeneous and relatively large spectral coverage, from
  3200 to 9500\AA.
\item The high spectral resolution and sampling for such a spectral coverage.
\item The wide metallicity range, although the present sample still
  needs some completion for extreme metallicities.
\end{enumerate}
We have presented some qualitative examples on the possible use of
STELIB for population synthesis and evolutionary models of
galaxies. Fundamental Plane studies of high redshift galaxies could be
another possible application for STELIB (Treu et al. \cite{Treu}).
In general, STELIB should be a useful tool for detailed
studies of galaxies at z\ltapprox 1, based on optical spectra at
intermediate resolution.

\acknowledgements
We would like to thanks J.-C. Mermilliod who provided us a file extracted from
the Lausanne photometric database. Many thanks to St\'epane Charlot for a careful 
reading of the manuscript. This research has made use of the SIMBAD database, 
operated at CDS, Strasbourg, France. We are grateful to the help of the staff of 
Roque de los Muchachos Observatory at La Palma and of Siding Spring Observatory 
in Australia, where these observations were conducted. Some examples shown in 
this paper come from observations collected at the European Southern Observatory, 
Chile (ESO N° 64.O-0439). Part of this work was supported by the French {\it 
Centre National de la Recherche Scientifique}, and by the French {\it Programme 
National Galaxies} (PNG). G. Bruzual acknowledges ample support from the 
Venezuelan Ministerio de Ciencia y Tecnolog\'\i a and FONACIT. G. Bruzual also 
thanks Observatoire Midi-Pyr\'en\'ees and the MENRT for their support during 
stays in Toulouse.

\appendix
\section{Synthetic photometry and photometric reliability of STELIB}
We present in this section a detailed comparison between the synthetic photometry 
derived for the STELIB library and the Lausanne database (Mermilliod et al. 
1997). Magnitudes for STELIB stars have been obtained using the flux calibrated 
spectra without any correction for dereddening or radial velocity. The 
photometric bands are UBVRI, with filter transmissions as close as possible to 
the Johnson filters commonly used in the Lausanne database. Table~\ref{tabfilt} 
summarizes the characteristics of the different filters. \\
Tables \ref{antab1} to \ref{antab4} provide with the UVBRI synthetic photometry 
for most of STELIB stars, together with the photoelectric photometry coming from 
the Lausanne database. Stars known to be variable or for which the spectrum is 
incomplete in a given filter have been discarded from this analysis. Almost all 
stars in this library have UBV Johnson magnitudes available, whereas R and I 
magnitudes are available only for 70\% of the whole sample. In addition, 
magnitudes in the R and I bands are given either in the Johnson system or in the 
Eggen or Cousins systems. The later are identified by a comment in the last
column of Tables \ref{antab1} to \ref{antab4}. When a filter band is missing in 
the STELIB spectra, the corresponding magnitude is given by "-" in the tables. 
A small extrapolation up to $\sim$100\AA\ is allowed in U and I when needed, 
towards the blue and the red edges of the filters. The following caveats apply in 
the comparison of photoelectric with synthetic magnitudes derived from our 
spectra:

\begin{itemize}
\item The effective filter transmissions in the different bands could be 
different between the photoelectric and the synthetic magnitudes.
\item The transmission of the Johnson U band has been set to zero in our 
calculations for wavelengths shorter than 3200\AA, which corresponds to the blue 
limit of the STELIB spectra.
\item Although the majority of stars in the Lausanne database are given in the 
Johnson system, STELIB spectra do not cover completely the standard I band. Thus, 
the effective transmission in the I band has been adapted to the red limit of our 
spectra in wavelength. When spectra extend up to $\sim$9800\AA, we use the 
broadest and redest version of the I band filter, identified by $I$ in
Table~\ref{tabfilt}, which is closer (although not identical) to a true Johnson 
filter. For spectra with red wavelegth limits bluer than $\sim$9800\AA, we used 
the I Cousins filter instead, identified by $I_{Cousins}$ in Table~\ref{tabfilt}. 
These stars are identified by an asterisk in Tables \ref{antab1} to \ref{antab4}.
\end{itemize}

\begin{table}
\centering
\caption[]{\label{tabfilt}Characteristics of filters used in Appendix A: the
effective wavelength $\lambda_{\rm eff}$ and the band width.}
\begin{flushleft}
\begin{tabular}{ccc}
\hline\noalign{\smallskip}
Filter       &  $\lambda_{\rm eff}$\,[\AA]  &  width\,[\AA]  \\
\noalign{\smallskip}
\hline\noalign{\smallskip}
$U$     &   3594  &   377 \\
$B$     &   4462  &   742 \\
$V$     &   5554  &   736 \\
$R$     &   6939  &  1507 \\
$I$     &   8548  &  1307 \\
$I_{Cousins}$ & 8060 & 924 \\
\noalign{\smallskip}
\hline
\end{tabular}
\end{flushleft}
\end{table}

\begin{table}
\centering
\caption[]{\label{tabdisp} Dispersion values in the comparison between synthetic 
magnitudes and photoelectric photometry from the Lausanne database}
\begin{flushleft}
\begin{tabular}{lrclr}
\hline\noalign{\smallskip}
   quantity &  rms & \hspace{0.5cm}   &   quantity &  rms \\
\noalign{\smallskip}
\hline\noalign{\smallskip}
$\Delta U$    &     0.145 &    &  $\Delta U-B$  &     0.156 \\
$\Delta B$    &     0.044 &    &  $\Delta B-V$  &     0.083 \\
$\Delta V$    &     0.072 &    &  $\Delta V-R$  &     0.135 \\
$\Delta R$    &     0.172 &    &  $\Delta R-I$  &     0.177 \\
$\Delta I$    &     0.249 &    &               &          \\
\noalign{\smallskip}
\hline
\end{tabular}
\end{flushleft}
\end{table}

\begin{figure}
\includegraphics[width=8cm,angle=0]{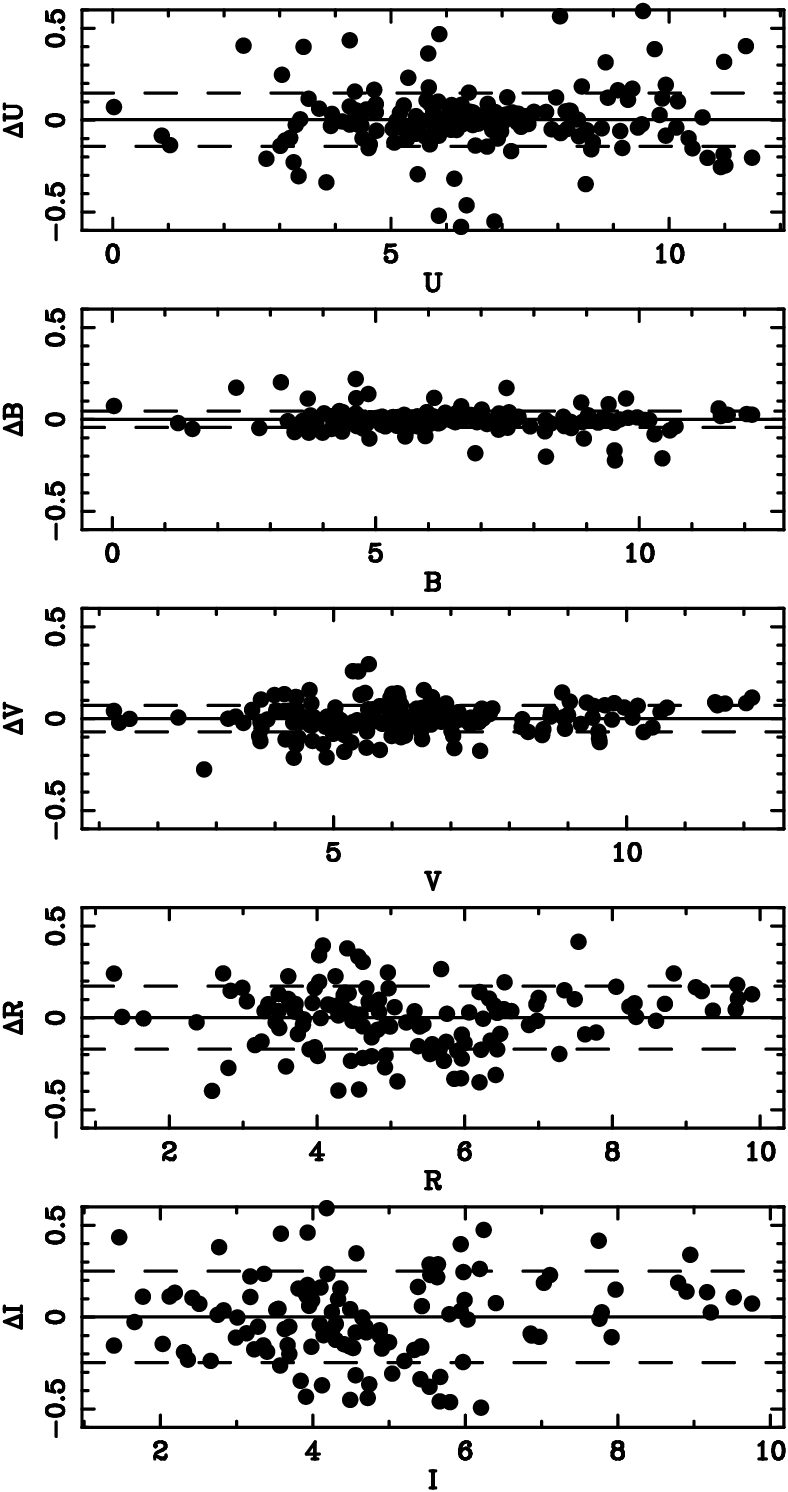}
\caption{Residuals of the comparison between synthetic photometry and
published photoelectric photometry: UBVRI magnitude residuals versus
Johnson magnitudes in the Lausanne database. Dashed lines correspond
to 1 $\sigma$ rms. 
} \label{figan1}
\end{figure}

\begin{figure}
\includegraphics[width=8cm,angle=0]{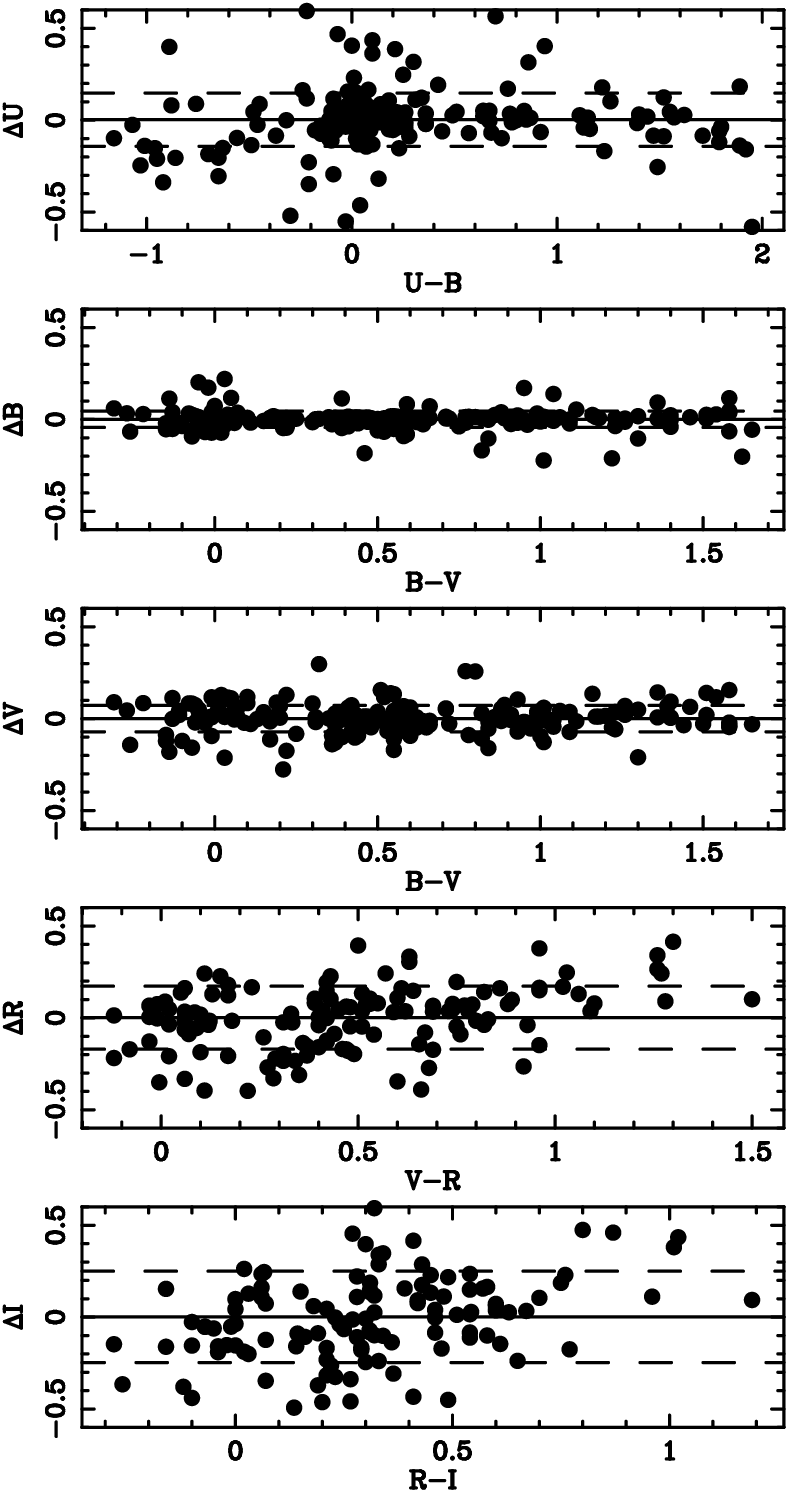}
\caption{Residuals of the comparison between synthetic photometry and
published photoelectric photometry: UBVRI magnitude residuals versus
Johnson colors in the Lausanne database. Dashed lines correspond
to 1 $\sigma$ rms. 
} \label{figan2}
\end{figure}

\begin{figure}
\includegraphics[width=8cm,angle=0]{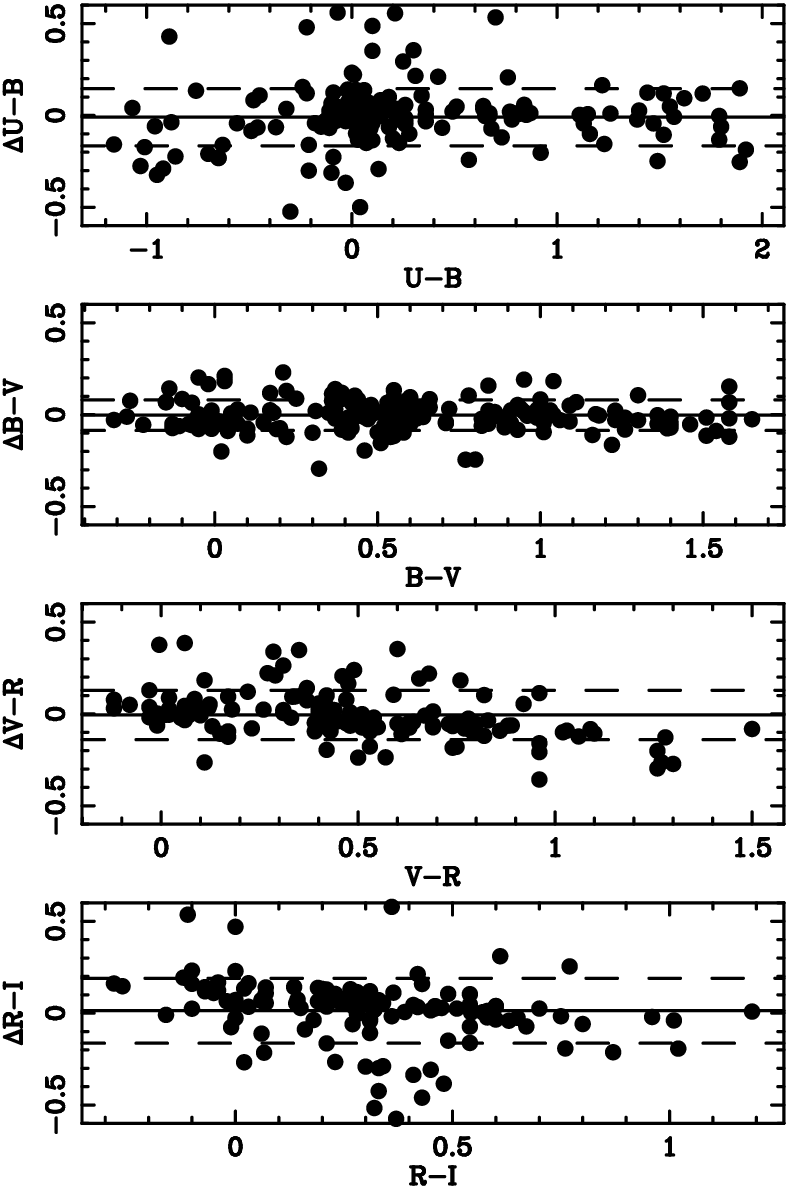}
\caption{Residuals of the comparison between synthetic photometry and
published photoelectric photometry: color residuals versus 
Johnson colors in the Lausanne database. Dashed lines correspond
to 1 $\sigma$ rms.
} \label{figan3}
\end{figure}

Figures \ref{figan1} to \ref{figan3} display the residuals of the comparison 
between synthetic photometry and the published Lausanne database. 
Table~\ref{tabdisp} summarizes the dispersion values obtained in the different 
filters and colors. As expected, the smallest dispersions correspond to the B and 
V magnitudes and B-V colors, for which we have the highest degree of confidence 
in the correspondance between filter bands. For these 2 filters, the rms
dispersion is quite consistent with the photometric accuracy derived from 
standard stars. The dispersion is much higher in the I band, as expected taking 
into account the inhomogeneities both in the photometric systems and the 
wavelegth coverage for the different objects. The situation in the U and R bands 
are intermediate. The wavelegth coverage could be responsible for the dispersion 
in U, (where the photometric systems are more consistent than in I), whereas the 
culprit in R is more likely the photometric system, but this point is difficult 
to assess. According to figures~\ref{figan1} to \ref{figan3}, there is no obvious 
color trend in the residuals neither in magnitude nor in color, except for the 
R-I and maybe in V-R in Figure~\ref{figan3}. This color trend is due to a 
residual difference between the filters used to compute synthetic magnitudes and 
the true Johnson filter. On the other hand, the relatively small dispersion in 
the R-I residuals as compared to the corresponding dispersion in R and I 
magnitudes (Figure~\ref{figan3}) indicates that the two photometric systems used 
are both internally consistent, but different from each other. 

\begin{table*}
\begin{flushleft}
\begin{tabular}{|l|rrrrr|rrrrr|r|}
\hline\noalign{\smallskip}
 &  \multicolumn{5}{c|}{photoelectric photometry }  &  \multicolumn{5}{c|}{synthetic photometry from spectra}  & \\
 Star & U & B & V & R & I & U & B & V & R & I & comment \\
\noalign{\smallskip}
\hline\noalign{\smallskip}
HD 2857    &   10.42  &   10.19  &    9.99  &    -     &    -     &   10.27  &   10.19  &   10.06  &    9.92  &    9.80 &                      \\ 
HD 6268    &    9.26  &    8.95  &    8.11  &    -     &    -     &    9.37  &    8.85  &    8.06  &    7.45  &    7.02 &                      \\ 
HD 15318   &    4.12  &    4.23  &    4.28  &    4.26  &    4.31  &    4.11  &    4.24  &    4.32  &    4.31  &    4.31 & *                    \\ 
HD 21581   &    9.74  &    9.53  &    8.71  &    7.97  &    7.43  &   10.13  &    9.36  &    8.60  &    8.05  &    7.64 &                      \\ 
HD 26630   &    5.75  &    5.11  &    4.15  &    3.36  &    2.82  &    5.77  &    5.08  &    4.13  &    3.43  &    3.12 & *                    \\ 
HD 30614   &    3.43  &    4.32  &    4.29  &    4.18  &    4.18  &    3.83  &    4.29  &    4.08  &    3.78  &    3.38 &                      \\ 
HD 30739   &    4.35  &    4.37  &    4.36  &    4.30  &    4.30  &    4.51  &    4.38  &    4.41  &    4.33  &    4.32 &                      \\ 
HD 32034   &    9.16  &    9.79  &    9.69  &    9.52  &    9.46  &    9.01  &    9.80  &    9.77  &    9.70  &    9.63 & *                    \\ 
HD 32923   &    5.71  &    5.57  &    4.92  &    4.65  &    4.42  &    5.80  &    5.57  &    4.87  &    4.38  &    4.48 & * Eggen              \\ 
HD 33579   &    9.08  &    9.32  &    9.13  &    8.90  &    8.75  &    9.24  &    9.33  &    9.22  &    9.07  &    8.95 & *                    \\ 
HD 34411   &    5.45  &    5.33  &    4.71  &    4.18  &    3.86  &    5.47  &    5.33  &    4.69  &    4.26  &    4.51 & *                    \\ 
HD 34816   &    3.01  &    4.02  &    4.29  &    4.41  &    4.69  &    2.87  &    4.05  &    4.33  &    4.42  &    4.60 &                      \\ 
HD 35497   &    1.03  &    1.52  &    1.65  &    1.66  &    1.76  &    0.89  &    1.47  &    1.65  &    1.66  &    1.79 & *                    \\ 
HD 36512   &    3.29  &    4.36  &    4.62  &    4.74  &    5.00  &    3.26  &    4.29  &    4.48  &    4.52  &    4.70 & *                    \\ 
HD 36673   &    3.04  &    2.79  &    2.58  &    2.36  &    2.15  &    3.29  &    2.74  &    2.30  &    1.96  &    1.98 & *                    \\ 
HD 37394   &    7.57  &    7.06  &    6.22  &    5.53  &    5.10  &    7.62  &    7.06  &    6.06  &    5.36  &    5.45 & *                    \\ 
HD 37828   &    8.86  &    8.00  &    6.87  &    5.94  &    5.27  &    9.18  &    -     &    -     &    5.90  &    5.36 &                      \\ 
HD 38247   &    9.94  &    8.23  &    6.61  &    -     &    -     &    9.86  &    8.03  &    -     &    4.44  &    3.20 &                      \\ 
HD 39587   &    5.06  &    4.99  &    4.40  &    3.89  &    3.58  &    4.94  &    4.99  &    4.40  &    3.93  &    3.79 &                      \\ 
HD 39866   &    7.18  &    6.92  &    6.62  &    -     &    -     &    7.22  &    6.90  &    6.70  &    6.47  &    6.50 & *                    \\ 
HD 39949   &    -     &    8.32  &    7.23  &    -     &    -     &    9.09  &    8.30  &    7.16  &    6.50  &    6.80 & *                    \\ 
HD 40111   &    3.84  &    4.76  &    4.82  &    4.76  &    4.87  &    3.50  &    4.71  &    4.80  &    4.70  &    4.33 &                      \\ 
HD 41636   &    8.23  &    7.39  &    6.35  &    -     &    -     &    8.28  &    7.38  &    6.34  &    5.65  &    4.98 &                      \\ 
HD 41667   &   10.10  &    9.54  &    8.53  &    -     &    -     &    -     &    9.32  &    8.40  &    7.75  &    7.28 &                      \\ 
HD 42454   &    9.34  &    8.58  &    7.35  &    6.39  &    5.70  &    9.51  &    8.54  &    7.29  &    6.54  &    -    & *                    \\ 
HD 43153   &    4.73  &    5.18  &    5.33  &    -     &    -     &    4.82  &    5.16  &    5.24  &    5.23  &    5.32 & *                    \\ 
HD 45829   &    9.83  &    8.21  &    6.63  &    5.54  &    4.78  &    9.86  &    8.14  &    6.58  &    5.58  &    5.07 & *                    \\ 
HD 46223   &    6.74  &    7.50  &    7.28  &    6.97  &    6.81  &    6.83  &    7.45  &    7.10  &    6.77  &    6.76 & *                    \\ 
HD 47129   &    5.23  &    6.11  &    6.06  &    5.97  &    5.91  &    5.31  &    6.23  &    6.17  &    6.00  &    6.22 & *                    \\ 
HD 47731   &    8.36  &    7.51  &    6.42  &    6.07  &    5.71  &    8.36  &    7.51  &    6.46  &    5.76  &    4.88 & Eggen                \\ 
HD 48329   &    5.86  &    4.39  &    2.99  &    2.03  &    1.42  &    5.78  &    4.35  &    2.99  &    2.19  &    1.33 &                      \\ 
HD 48682   &    5.85  &    5.79  &    5.24  &    -     &    -     &    5.95  &    5.75  &    5.07  &    4.56  &    4.34 &                      \\ 
HD 49933   &    6.05  &    6.15  &    5.76  &    -     &    -     &    6.10  &    6.13  &    5.66  &    5.26  &    5.13 & *                    \\ 
HD 52005   &    9.12  &    7.33  &    5.68  &    4.42  &    3.59  &    9.06  &    7.27  &    5.65  &    4.69  &    -    & *                    \\ 
HD 53929   &    5.51  &    5.97  &    6.10  &    -     &    -     &    5.48  &    6.01  &    6.21  &    6.21  &    6.28 &                      \\ 
HD 54719   &    7.07  &    5.67  &    4.41  &    3.45  &    2.82  &    7.10  &    5.67  &    4.43  &    3.83  &    -    &                      \\ 
HD 58551   &    -     &    7.00  &    6.54  &    6.12  &    5.81  &    6.87  &    6.99  &    6.54  &    6.31  &    -    &                      \\ 
HD 59881   &    5.64  &    5.46  &    5.24  &    -     &    -     &    5.75  &    5.47  &    5.37  &    5.31  &    5.72 & *                    \\ 
HD 60178   &    -     &    -     &    -     &    -     &    -     &    2.11  &    2.03  &    1.50  &    0.63  &    -    &                      \\ 
HD 60778   &    9.44  &    9.22  &    9.11  &    -     &    -     &    9.40  &    9.20  &    9.08  &    8.94  &    9.13 &                      \\ 
HD 61064   &    5.67  &    5.57  &    5.13  &    -     &    -     &    6.03  &    5.58  &    5.06  &    4.78  &    5.16 &                      \\ 
HD 63077   &    5.87  &    5.94  &    5.36  &    -     &    -     &    6.34  &    5.85  &    5.37  &    4.96  &    4.68 &                      \\ 
HD 64090   &    8.79  &    8.92  &    8.30  &    7.75  &    7.34  &    8.75  &    8.91  &    8.31  &    7.83  &    7.82 & *                    \\ 
HD 65583   &    7.89  &    7.71  &    7.00  &    6.40  &    5.98  &    7.84  &    7.72  &    7.06  &    6.51  &    6.12 &                      \\ 
HD 67767   &    6.98  &    6.54  &    5.72  &    5.41  &    5.14  &    6.92  &    6.55  &    5.75  &    5.18  &    4.87 & * Eggen              \\ 
HD 69897   &    5.54  &    5.60  &    5.14  &    -     &    -     &    5.54  &    5.60  &    5.13  &    4.72  &    4.52 &                      \\ 
HD 72184   &    8.17  &    7.01  &    5.90  &    5.43  &    5.03  &    8.12  &    7.06  &    5.89  &    5.25  &    5.69 & * Eggen              \\ 
HD 72324   &    8.25  &    7.38  &    6.35  &    5.93  &    5.58  &    8.26  &    7.38  &    6.33  &    5.81  &    -    & *                    \\ 
HD 74739   &    5.82  &    5.04  &    4.03  &    3.28  &    2.79  &    5.81  &    5.04  &    4.05  &    3.48  &    3.76 & *                    \\ 
HD 75732   &    7.46  &    6.81  &    5.95  &    5.66  &    5.40  &    7.48  &    6.83  &    5.96  &    5.34  &    5.00 & *                    \\ 
HD 76151   &    6.88  &    6.66  &    6.00  &    5.64  &    5.31  &    6.92  &    6.67  &    5.97  &    5.50  &    5.66 & * Cousins            \\ 
HD 76943   &    4.44  &    4.40  &    3.97  &    3.57  &    3.35  &    4.45  &    4.40  &    3.87  &    3.41  &    3.15 & *                    \\ 
\hline\noalign{\smallskip}
\end{tabular}
\end{flushleft}
\caption[]{ \label{antab1} Photometry of STELIB stars from Lausanne database and synthetic photometry from spectra}
\end{table*}
\begin{table*}
\begin{flushleft}
\begin{tabular}{|l|rrrrr|rrrrr|r|}
\hline\noalign{\smallskip}
 &  \multicolumn{5}{c|}{photoelectric photometry}  &  \multicolumn{5}{c|}{synthetic photometry from spectra}  & \\
 Star & U & B & V & R & I & U & B & V & R & I  & comment \\
\noalign{\smallskip}
\hline\noalign{\smallskip}
HD 77350   &    5.29  &    5.40  &    5.44  &    5.42  &    5.46  &    5.19  &    5.37  &    5.43  &    5.38  &    5.36 &                      \\ 
HD 77729   &   10.60  &    9.03  &    7.63  &    6.87  &    6.29  &   10.62  &    9.05  &    7.72  &    6.78  &    6.25 & * Eggen              \\ 
HD 78418   &    6.82  &    6.62  &    5.96  &    5.67  &    5.44  &    6.77  &    6.69  &    5.95  &    5.45  &    5.18 & * Eggen              \\ 
HD 79158   &    4.70  &    5.18  &    5.32  &    -     &    -     &    4.75  &    5.14  &    5.14  &    4.97  &    4.86 &                      \\ 
HD 79452   &    7.20  &    6.84  &    6.00  &    -     &    -     &    7.24  &    6.84  &    5.98  &    5.37  &    4.90 &                      \\ 
HD 80081   &    3.92  &    3.87  &    3.81  &    3.69  &    3.66  &    3.89  &    3.85  &    3.81  &    3.65  &    3.52 & *                    \\ 
HD 81192   &    8.05  &    7.48  &    6.53  &    5.79  &    5.25  &    7.98  &    7.65  &    6.51  &    5.84  &    5.33 &                      \\ 
HD 81809   &    6.14  &    6.01  &    5.37  &    5.00  &    4.64  &    5.82  &    5.98  &    5.36  &    4.85  &    4.56 & * Cousins            \\ 
HD 82210   &    5.67  &    5.34  &    4.57  &    3.91  &    3.50  &    -     &    5.32  &    -     &    3.52  &    3.13 & *                    \\ 
HD 83632   &   10.93  &    9.44  &    8.05  &    7.03  &    6.28  &   10.68  &    9.43  &    8.12  &    7.20  &    6.53 &                      \\ 
HD 84937   &    8.50  &    8.71  &    8.32  &    7.92  &    7.64  &    8.15  &    8.66  &    8.36  &    7.92  &    7.59 &                      \\ 
HD 85235   &    4.70  &    4.62  &    4.59  &    4.49  &    4.49  &    4.87  &    4.84  &    4.60  &    4.51  &    4.59 &                      \\ 
HD 86728   &    6.28  &    6.02  &    5.37  &    -     &    -     &    6.26  &    6.03  &    5.34  &    4.79  &    4.41 &                      \\ 
HD 87141   &    6.27  &    6.23  &    5.75  &    -     &    -     &    6.35  &    6.24  &    5.70  &    5.26  &    4.95 &                      \\ 
HD 87696   &    4.73  &    4.66  &    4.48  &    4.30  &    4.23  &    4.77  &    4.68  &    4.49  &    4.28  &    4.17 &                      \\ 
HD 87737   &    3.25  &    3.46  &    3.49  &    3.40  &    3.38  &    3.02  &    3.39  &    3.47  &    3.35  &    3.25 &                      \\ 
HD 87822   &    6.69  &    6.68  &    6.24  &    -     &    -     &    6.70  &    6.69  &    6.22  &    5.80  &    5.53 &                      \\ 
HD 87901   &    0.88  &    1.25  &    1.36  &    1.39  &    1.49  &    0.80  &    1.23  &    1.40  &    1.40  &    1.40 &                      \\ 
HD 88355   &    6.92  &    6.90  &    6.44  &    6.03  &    5.76  &    6.82  &    6.90  &    6.45  &    6.06  &    5.81 &                      \\ 
HD 88609   &    9.94  &    9.52  &    8.59  &    7.79  &    7.16  &   10.13  &    9.50  &    8.52  &    7.77  &    7.25 &                      \\ 
HD 89025   &    3.94  &    3.75  &    3.44  &    3.13  &    2.94  &    3.97  &    3.75  &    3.42  &    3.11  &    2.91 &                      \\ 
HD 89254   &    5.70  &    5.60  &    5.28  &    -     &    -     &    5.57  &    5.60  &    5.58  &    5.52  &    5.56 &                      \\ 
HD 89758   &    6.52  &    4.63  &    3.05  &    1.77  &    0.81  &    6.38  &    4.75  &    3.01  &    1.86  &    0.98 &                      \\ 
HD 90277   &    5.16  &    4.99  &    4.74  &    4.48  &    4.34  &    5.20  &    4.99  &    4.66  &    4.37  &    4.24 & *                    \\ 
HD 90508   &    7.09  &    7.04  &    6.44  &    5.97  &    5.67  &    7.21  &    7.04  &    6.35  &    5.80  &    5.49 & *                    \\ 
HD 90537   &    5.76  &    5.12  &    4.21  &    3.52  &    3.06  &    5.81  &    5.13  &    4.20  &    3.59  &    3.16 &                      \\ 
HD 91316   &    2.76  &    3.71  &    3.85  &    3.89  &    4.05  &    2.55  &    3.82  &    -     &    -     &    4.26 &                      \\ 
HD 92769   &    5.77  &    5.68  &    5.51  &    -     &    -     &    5.77  &    5.69  &    5.49  &    5.28  &    5.11 &                      \\ 
HD 93250   &    -     &    -     &    -     &    -     &    -     &    6.54  &    7.53  &    7.41  &    7.15  &    7.03 &                      \\ 
HD 93430   &    -     &    -     &    -     &    -     &    -     &   10.01  &   10.05  &    9.57  &    9.16  &    8.85 &                      \\ 
HD 93765   &    6.40  &    6.43  &    6.06  &    -     &    -     &    6.43  &    6.43  &    6.03  &    5.68  &    5.46 &                      \\ 
HD 94028   &    8.52  &    8.70  &    8.23  &    7.76  &    7.46  &    8.47  &    8.69  &    8.24  &    7.82  &    7.51 &                      \\ 
HD 94247   &    7.97  &    6.45  &    5.09  &    4.49  &    4.00  &    8.09  &    6.45  &    5.10  &    4.14  &    3.61 & * Eggen              \\ 
HD 94264   &    5.78  &    4.86  &    3.82  &    2.99  &    2.45  &    5.72  &    5.00  &    3.78  &    2.98  &    2.40 &                      \\ 
HD 95345   &    7.11  &    6.00  &    4.84  &    4.25  &    3.71  &    7.14  &    6.02  &    4.98  &    4.28  &    3.80 & Cousins              \\ 
HD 95418   &    2.35  &    2.35  &    2.37  &    2.31  &    2.35  &    2.76  &    2.52  &    2.38  &    2.29  &    2.22 &                      \\ 
HD 95735   &   10.13  &    9.00  &    7.49  &    5.99  &    4.80  &   10.09  &    9.00  &    7.51  &    6.09  &    4.95 &                      \\ 
HD 97633   &    3.37  &    3.33  &    3.34  &    3.35  &    3.35  &    3.37  &    3.32  &    3.35  &    3.42  &    3.26 &                      \\ 
HD 97916   &    9.51  &    9.63  &    9.21  &    8.79  &    8.48  &    9.49  &    9.63  &    9.28  &    8.94  &    8.73 &                      \\ 
HD 98262   &    6.43  &    4.88  &    3.48  &    2.42  &    1.72  &    6.48  &    4.88  &    3.49  &    2.55  &    1.89 &                      \\ 
HD 98839   &    6.79  &    5.99  &    4.99  &    4.24  &    3.78  &    6.79  &    5.98  &    4.90  &    4.19  &    3.76 &                      \\ 
HD 99028   &    4.42  &    4.35  &    3.94  &    3.55  &    3.34  &    4.39  &    4.37  &    3.97  &    3.63  &    3.45 &                      \\ 
HD 99747   &    6.14  &    6.22  &    5.86  &    -     &    -     &    6.15  &    6.20  &    5.77  &    5.34  &    5.04 &                      \\ 
HD 100006  &    7.41  &    6.59  &    5.53  &    5.04  &    4.68  &    7.43  &    6.60  &    5.57  &    4.84  &    4.43 & * Eggen              \\ 
HD 100889  &    4.47  &    4.62  &    4.70  &    4.69  &    4.76  &    4.44  &    4.65  &    4.78  &    4.78  &    4.77 & *                    \\ 
HD 101501  &    6.30  &    6.04  &    5.32  &    4.70  &    4.39  &    6.29  &    6.04  &    5.29  &    4.74  &    4.37 &                      \\ 
HD 101606  &    6.08  &    6.19  &    5.75  &    5.33  &    5.04  &    6.08  &    6.17  &    5.66  &    5.20  &    4.92 & *                    \\ 
HD 102212  &    7.34  &    5.54  &    4.03  &    2.77  &    1.76  &    7.30  &    5.57  &    4.17  &    3.11  &    2.20 &                      \\ 
HD 102224  &    6.04  &    4.89  &    3.71  &    2.83  &    2.23  &    6.05  &    4.90  &    3.72  &    2.91  &    2.33 &                      \\ 
HD 102634  &    6.73  &    6.66  &    6.14  &    -     &    -     &    6.59  &    6.67  &    6.26  &    5.93  &    5.74 &                      \\ 
HD 102870  &    4.26  &    4.16  &    3.61  &    3.18  &    2.90  &    4.70  &    4.11  &    3.65  &    3.28  &    3.07 & *                    \\ 
HD 102870  &    4.26  &    4.16  &    3.61  &    3.18  &    2.90  &    4.24  &    4.18  &    3.74  &    3.41  &    3.18 & *                    \\ 
\hline\noalign{\smallskip}
\end{tabular}
\end{flushleft}
\caption[]{ \label{antab2} Photometry of STELIB stars from Lausanne database and 
synthetic photometry from spectra (cont'd)}
\end{table*}
\begin{table*}
\begin{flushleft}
\begin{tabular}{|l|rrrrr|rrrrr|r|}
\hline\noalign{\smallskip}
 &  \multicolumn{5}{c|}{photoelectric photometry}  &  \multicolumn{5}{c|}{synthetic photometry from spectra}  & \\
 Star & U & B & V & R & I & U & B & V & R & I  & comment \\
\noalign{\smallskip}
\hline\noalign{\smallskip}
HD 104893  &   11.38  &   10.44  &    9.22  &    -     &    -     &   11.78  &   10.23  &    9.17  &    8.39  &    7.78 &                      \\ 
HD 105546  &    -     &    -     &    -     &    -     &    -     &    9.28  &    9.04  &    8.22  &    7.54  &    7.04 &                      \\ 
HD 106038  &    -     &    -     &    -     &    -     &    -     &   10.15  &   10.26  &    9.67  &    9.12  &    8.75 &                      \\ 
HD 107213  &    7.01  &    6.93  &    6.40  &    -     &    -     &    7.00  &    6.94  &    6.40  &    5.97  &    5.70 &                      \\ 
HD 108177  &    9.88  &   10.10  &    9.67  &    9.22  &    8.90  &   10.00  &   10.10  &    9.68  &    9.27  &    8.99 &                      \\ 
HD 109995  &    7.79  &    7.66  &    7.61  &    -     &    -     &    7.84  &    7.66  &    7.63  &    7.51  &    7.40 &                      \\ 
HD 110897  &    6.47  &    6.51  &    5.96  &    5.42  &    5.13  &    6.44  &    6.50  &    5.85  &    5.33  &    5.03 & *                    \\ 
HD 111028  &    7.43  &    6.65  &    5.66  &    5.20  &    4.87  &    7.43  &    6.67  &    5.70  &    5.03  &    4.69 & * Eggen              \\ 
HD 112033  &    -     &    -     &    -     &    -     &    -     &   10.76  &   10.51  &    9.85  &    9.30  &    8.91 &                      \\ 
HD 113022  &    6.63  &    6.63  &    6.20  &    6.21  &    6.07  &    6.64  &    6.64  &    6.22  &    5.85  &    5.64 & * Eggen              \\ 
HD 113139  &    5.31  &    5.30  &    4.93  &    4.56  &    4.35  &    5.54  &    5.31  &    4.80  &    4.36  &    4.09 & *                    \\ 
HD 113226  &    4.49  &    3.76  &    2.83  &    2.19  &    1.74  &    4.39  &    3.78  &    2.93  &    2.34  &    1.93 &                      \\ 
HD 113337  &    6.48  &    6.49  &    6.07  &    -     &    -     &    6.52  &    6.49  &    6.00  &    5.62  &    5.38 &                      \\ 
HD 114330  &    4.37  &    4.37  &    4.38  &    4.33  &    4.33  &    4.42  &    4.41  &    4.50  &    4.47  &    4.49 & *                    \\ 
HD 115383  &    5.89  &    5.79  &    5.21  &    4.88  &    4.57  &    5.90  &    5.81  &    5.28  &    4.85  &    4.56 & Cousins              \\ 
HD 116842  &    4.26  &    4.18  &    4.01  &    3.84  &    3.77  &    4.34  &    4.18  &    3.90  &    3.63  &    3.48 & *                    \\ 
HD 117176  &    5.94  &    5.68  &    4.97  &    4.36  &    3.97  &    5.90  &    5.69  &    5.02  &    4.52  &    4.19 &                      \\ 
HD 120136  &    5.03  &    4.98  &    4.50  &    4.09  &    3.85  &    4.98  &    4.99  &    4.51  &    4.11  &    3.87 &                      \\ 
HD 122408  &    4.47  &    4.35  &    4.25  &    4.10  &    4.04  &    4.51  &    4.36  &    4.37  &    4.33  &    4.26 &                      \\ 
HD 122563  &    7.47  &    7.11  &    6.20  &    5.38  &    4.80  &    7.45  &    7.09  &    6.22  &    5.52  &    5.03 &                      \\ 
HD 123299  &    3.52  &    3.61  &    3.66  &    3.69  &    3.76  &    3.64  &    3.60  &    3.71  &    3.76  &    3.77 &                      \\ 
HD 124425  &    6.40  &    6.38  &    5.91  &    -     &    -     &    6.55  &    6.39  &    5.92  &    5.55  &    5.32 &                      \\ 
HD 124570  &    6.18  &    6.09  &    5.55  &    -     &    -     &    6.13  &    6.11  &    5.69  &    5.31  &    5.05 &                      \\ 
HD 124850  &    4.62  &    4.59  &    4.08  &    3.58  &    3.31  &    4.49  &    4.59  &    4.24  &    3.97  &    3.83 &                      \\ 
HD 125560  &    7.46  &    6.07  &    4.84  &    3.95  &    3.35  &    7.44  &    6.08  &    4.88  &    4.05  &    3.47 &                      \\ 
HD 126141  &    6.57  &    6.60  &    6.23  &    -     &    -     &    6.55  &    6.61  &    6.24  &    5.90  &    5.64 &                      \\ 
HD 126660  &    4.56  &    4.55  &    4.05  &    3.63  &    3.38  &    4.62  &    4.55  &    4.04  &    3.63  &    3.38 &                      \\ 
HD 127665  &    6.32  &    4.88  &    3.58  &    2.66  &    2.01  &    6.34  &    4.78  &    3.37  &    2.40  &    1.83 & *                    \\ 
HD 128167  &    4.74  &    4.82  &    4.46  &    4.12  &    3.93  &    4.68  &    4.79  &    4.32  &    3.89  &    3.62 & *                    \\ 
HD 130109  &    3.71  &    3.73  &    3.74  &    3.67  &    3.69  &    3.77  &    3.66  &    3.65  &    3.58  &    3.60 &                      \\ 
HD 130948  &    6.43  &    6.42  &    5.86  &    5.80  &    5.60  &    6.46  &    6.44  &    5.91  &    5.47  &    5.20 & * Eggen              \\ 
HD 131156  &    5.61  &    5.33  &    4.56  &    3.93  &    3.50  &    5.52  &    5.34  &    4.82  &    4.26  &    3.74 &                      \\ 
HD 132142  &    8.90  &    8.56  &    7.78  &    7.11  &    6.66  &    9.02  &    8.57  &    7.69  &    7.03  &    6.95 &                      \\ 
HD 134083  &    5.33  &    5.36  &    4.93  &    4.53  &    4.32  &    5.29  &    5.36  &    4.90  &    4.49  &    4.21 &                      \\ 
HD 134169  &    8.14  &    8.22  &    7.69  &    -     &    -     &    8.19  &    8.22  &    7.69  &    7.21  &    6.93 &                      \\ 
HD 135722  &    5.10  &    4.43  &    3.48  &    2.75  &    2.24  &    5.10  &    4.43  &    3.46  &    2.79  &    2.31 &                      \\ 
HD 136512  &    7.29  &    6.52  &    5.51  &    -     &    -     &    7.32  &    6.53  &    5.57  &    4.89  &    4.40 &                      \\ 
HD 137759  &    5.68  &    4.46  &    3.29  &    2.51  &    1.91  &    5.86  &    4.47  &    3.30  &    2.55  &    2.04 &                      \\ 
HD 138290  &    6.87  &    6.93  &    6.56  &    -     &    -     &    6.89  &    6.94  &    6.59  &    6.26  &    6.04 &                      \\ 
HD 139641  &    6.64  &    6.15  &    5.26  &    -     &    -     &    6.67  &    6.16  &    5.33  &    5.64  &    4.48 &                      \\ 
HD 139669  &    8.43  &    6.54  &    4.96  &    3.93  &    3.06  &    8.61  &    6.58  &    5.12  &    4.18  &    3.58 & Eggen                \\ 
HD 139798  &    6.09  &    6.11  &    5.76  &    5.43  &    5.25  &    6.17  &    6.12  &    5.76  &    5.45  &    5.37 &                      \\ 
HD 141004  &    5.13  &    5.03  &    4.43  &    3.92  &    3.60  &    5.11  &    5.04  &    4.49  &    4.05  &    3.78 &                      \\ 
HD 141714  &    5.78  &    5.42  &    4.62  &    3.99  &    3.57  &    5.76  &    5.43  &    4.88  &    4.29  &    3.72 &                      \\ 
HD 142373  &    5.19  &    5.18  &    4.62  &    4.14  &    3.82  &    5.15  &    5.17  &    4.58  &    4.09  &    3.78 &                      \\ 
HD 144206  &    4.32  &    4.64  &    4.74  &    4.72  &    4.82  &    4.32  &    4.60  &    4.62  &    4.51  &    4.44 & *                    \\ 
HD 145675  &    8.20  &    7.53  &    6.65  &    -     &    -     &    8.25  &    7.57  &    6.70  &    6.07  &    5.72 &                      \\ 
HD 145976  &    6.85  &    6.90  &    6.52  &    -     &    -     &    6.90  &    6.91  &    6.55  &    6.21  &    5.97 &                      \\ 
HD 146051  &    6.26  &    4.31  &    2.73  &    1.46  &    0.44  &    5.68  &    4.36  &    2.71  &    1.70  &    0.94 &                      \\ 
HD 147394  &    3.19  &    3.75  &    3.90  &    3.98  &    4.14  &    3.09  &    3.70  &    3.78  &    3.81  &    4.04 &                      \\ 
HD 148513  &    8.63  &    6.84  &    5.38  &    4.56  &    4.02  &    8.51  &    6.85  &    5.44  &    4.52  &    4.00 & Eggen                \\ 
HD 149121  &    5.44  &    5.59  &    5.64  &    -     &    -     &    5.37  &    5.56  &    5.69  &    5.65  &    5.59 &                      \\ 
\hline\noalign{\smallskip}
\end{tabular}
\end{flushleft}
\caption[]{ \label{antab3} Photometry of STELIB stars from Lausanne database and 
synthetic photometry from spectra (cont'd)}
\end{table*}
\begin{table*}
\begin{flushleft}
\begin{tabular}{|l|rrrrr|rrrrr|r|}
\hline\noalign{\smallskip}
 &  \multicolumn{5}{c|}{photoelectric photometry}  &  \multicolumn{5}{c|}{synthetic photometry from spectra}  & \\
 Star & U & B & V & R & I & U & B & V & R & I  & comment \\
\noalign{\smallskip}
\hline\noalign{\smallskip}
HD 150275  &    8.04  &    7.34  &    6.35  &    -     &    -     &    8.61  &    7.37  &    6.36  &    5.66  &    5.23 &                      \\ 
HD 151044  &    7.03  &    7.01  &    6.48  &    6.04  &    5.78  &    7.04  &    7.01  &    6.41  &    5.95  &    -    &                      \\ 
HD 151217  &    8.60  &    6.68  &    5.14  &    -     &    -     &    8.44  &    6.71  &    5.26  &    4.15  &    2.30 &                      \\ 
HD 154733  &    8.38  &    6.86  &    5.56  &    4.90  &    4.43  &    8.29  &    6.88  &    5.61  &    4.76  &    4.32 & * Eggen              \\ 
HD 155646  &    7.17  &    7.13  &    6.64  &    -     &    -     &    -     &    -     &    6.63  &    6.27  &    -    &                      \\ 
HD 156283  &    6.27  &    4.60  &    3.16  &    2.20  &    1.48  &    -     &    -     &    3.12  &    2.05  &    -    &                      \\ 
HD 157089  &    7.53  &    7.55  &    6.97  &    6.45  &    6.10  &    -     &    -     &    6.97  &    6.52  &    -    &                      \\ 
HD 157214  &    6.08  &    6.01  &    5.39  &    4.88  &    4.54  &    6.03  &    6.00  &    5.34  &    4.83  &    4.50 &                      \\ 
HD 157373  &    6.72  &    6.82  &    6.41  &    -     &    -     &    -     &    -     &    6.41  &    6.06  &    5.98 & *                    \\ 
HD 157856  &    6.86  &    6.89  &    6.43  &    -     &    -     &    6.31  &    6.71  &    6.44  &    -     &    -    &                      \\ 
HD 157881  &   10.16  &    8.90  &    7.54  &    6.24  &    5.44  &   10.26  &    8.99  &    7.68  &    6.65  &    5.98 &                      \\ 
HD 157999  &    7.43  &    5.83  &    4.33  &    3.23  &    2.46  &    -     &    -     &    4.30  &    3.31  &    2.35 & *                    \\ 
HD 159181  &    4.40  &    3.77  &    2.80  &    2.12  &    1.64  &    -     &    -     &    2.75  &    1.85  &    1.81 &                      \\ 
HD 159332  &    6.15  &    6.15  &    5.66  &    -     &    -     &    -     &    -     &    5.66  &    8.62  &    -    &                      \\ 
HD 160693  &    8.92  &    8.95  &    8.37  &    7.89  &    7.55  &    -     &    -     &    8.37  &    -     &    -    &                      \\ 
HD 161817  &    7.27  &    7.13  &    6.98  &    6.86  &    6.71  &    7.26  &    7.12  &    7.02  &    6.84  &    6.69 & Cousins              \\ 
HD 163993  &    5.32  &    4.64  &    3.70  &    3.01  &    2.55  &    5.25  &    4.64  &    3.67  &    3.04  &    2.61 &                      \\ 
HD 164349  &    7.16  &    5.93  &    4.67  &    3.81  &    3.24  &    6.99  &    5.92  &    4.74  &    3.97  &    3.46 &                      \\ 
HD 164353  &    3.34  &    3.99  &    3.97  &    3.91  &    3.88  &    3.04  &    3.92  &    4.10  &    4.07  &    4.07 & *                    \\ 
HD 165908  &    5.48  &    5.57  &    5.05  &    4.57  &    4.23  &    5.19  &    5.50  &    5.04  &    4.63  &    4.64 & *                    \\ 
HD 166285  &    6.08  &    6.12  &    5.68  &    -     &    -     &    -     &    -     &    5.68  &    5.35  &    5.33 & *                    \\ 
HD 166620  &    7.87  &    7.28  &    6.40  &    5.63  &    5.14  &    -     &    -     &    6.39  &    5.70  &    5.42 & *                    \\ 
HD 167858  &    6.95  &    6.93  &    6.62  &    -     &    -     &    -     &    -     &    6.63  &    6.35  &    5.63 &                      \\ 
HD 172167  &    0.02  &    0.03  &    0.03  &    0.07  &    0.10  &    0.09  &    0.10  &    -     &    -     &    -    &                      \\ 
HD 173780  &    7.26  &    6.03  &    4.83  &    3.95  &    3.34  &    -     &    -     &    4.85  &    -     &    -    &                      \\ 
HD 173880  &    4.56  &    4.49  &    4.36  &    4.19  &    4.13  &    -     &    -     &    4.36  &    4.31  &    4.43 & *                    \\ 
HD 175305  &    8.08  &    7.93  &    7.18  &    6.53  &    6.04  &    -     &    7.89  &    -     &    -     &    -    & *                    \\ 
HD 175640  &    5.86  &    6.16  &    6.21  &    -     &    -     &    5.34  &    6.16  &    6.21  &    6.15  &    6.18 &                      \\ 
HD 176437  &    3.10  &    3.20  &    3.25  &    3.28  &    3.29  &    2.99  &    3.40  &    3.25  &    3.15  &    3.30 & *                    \\ 
HD 181470  &    6.17  &    6.25  &    6.26  &    -     &    -     &    -     &    -     &    6.27  &    6.18  &    6.39 & *                    \\ 
HD 182101  &    6.75  &    6.78  &    6.33  &    5.94  &    5.64  &    -     &    -     &    6.34  &    6.04  &    6.10 & *                    \\ 
HD 182490  &    6.36  &    6.32  &    6.25  &    6.19  &    6.17  &    5.90  &    6.36  &    6.26  &    6.19  &    6.49 & *                    \\ 
HD 184960  &    6.19  &    6.20  &    5.73  &    5.27  &    4.99  &    -     &    -     &    5.74  &    -     &    -    &                      \\ 
HD 185657  &    8.21  &    7.46  &    6.46  &    -     &    -     &    -     &    -     &    6.45  &    -     &    -    &                      \\ 
HD 188209  &    4.60  &    5.56  &    5.63  &    5.53  &    5.65  &    4.45  &    5.47  &    5.47  &    5.34  &    5.33 &                      \\ 
HD 188510  &    9.31  &    9.42  &    8.83  &    8.26  &    7.89  &    -     &    9.50  &    8.84  &    8.50  &    8.77 &                      \\ 
HD 195725  &    4.57  &    4.41  &    4.21  &    4.04  &    3.95  &    -     &    -     &    4.22  &    -     &    -    &                      \\ 
HD 195986  &    5.91  &    6.48  &    6.59  &    -     &    -     &    -     &    -     &    6.61  &    -     &    -    &                      \\ 
HD 197345  &    1.11  &    1.34  &    1.25  &    1.14  &    1.04  &    -     &    -     &    1.23  &    1.38  &    -    &                      \\ 
HD 268623  &   10.69  &   11.55  &   11.61  &    -     &    -     &   10.49  &   11.57  &   11.68  &   11.70  &   11.76 &                      \\ 
HD 268749  &   11.49  &   12.14  &   12.10  &    -     &    -     &   11.29  &   12.17  &   12.21  &   12.17  &   12.18 &                      \\ 
HD 268819  &   10.97  &   10.58  &   10.08  &    -     &    -     &    -     &   10.52  &   10.12  &    9.83  &    9.61 &                      \\ 
HD 269697  &   10.99  &   10.69  &   10.28  &    -     &    -     &   11.31  &   10.65  &   10.34  &   10.12  &    9.93 &                      \\ 
HD 269698  &   11.01  &   12.04  &   12.26  &    -     &    -     &   10.76  &   12.07  &   12.34  &   12.39  &   12.52 &                      \\ 
HD 269781  &    -     &    9.96  &    9.89  &    9.76  &    9.69  &    9.21  &    9.97  &    9.95  &    9.89  &    9.82 & *                    \\ 
HD 271163  &   10.98  &   11.68  &   11.75  &    -     &    -     &   10.80  &   11.70  &   11.83  &   11.84  &   11.91 &                      \\ 
HD 271182  &   10.74  &   10.29  &    9.70  &    9.17  &    8.86  &    -     &   10.21  &    9.63  &    9.27  &    9.06 & *                    \\ 
HD 338529  &    9.53  &    9.75  &    9.36  &    8.95  &    8.62  &   10.12  &    9.86  &    9.36  &    8.99  &    9.02 & *                    \\ 
Feige110   &   10.35  &   11.51  &   11.82  &    -     &    -     &   10.25  &   11.57  &   11.91  &   12.04  &   12.27 &                      \\ 
G319       &    -     &    -     &    -     &    -     &    -     &   13.17  &   12.71  &   12.58  &   12.47  &   12.35 &                      \\ 
LTT4364    &    -     &    -     &    -     &    -     &    -     &   10.72  &   11.64  &   11.49  &   11.39  &   11.30 &                      \\ 
\hline\noalign{\smallskip}
\end{tabular}
\end{flushleft}
\caption[]{ \label{antab4} Photometry of STELIB stars from Lausanne database and 
synthetic photometry from spectra (cont'd)}
\end{table*}

%\listofobjects
\end{document}